\documentclass{article}

\usepackage{arxiv}

\usepackage[utf8]{inputenc} 
\usepackage[T1]{fontenc}    
\usepackage{lmodern}       
\usepackage{hyperref}       
\usepackage{url}            
\usepackage{booktabs}       
\usepackage{amsfonts}       
\usepackage{microtype}      
\usepackage{lipsum}
\usepackage{graphicx}
\usepackage{caption} 
\usepackage{natbib}
\usepackage{amsmath}  
\usepackage{placeins} 

 \title{Generalizations of Langbein's Formula under Non-Stationarity, Mixed Populations, and Over- or Under-Dispersion in the Number of Exceedances}

\author{
 Francesco Dell'Aira \\
  Department of Civil, Construction, \\ and Environmental Engineering\\
  University of Memphis\\
  Memphis, TN 38152 \\
  \texttt{fdllaira@memphis.edu} \\
   \And
 \And
 Antonino Cancelliere \\
  Department of Civil \\ Engineering and Architecture \\
  University of Catania \\
  Catania, Italy \\
  \texttt{antonino.cancelliere@unict.it} \\
  \And
 Claudio I. Meier \\
  Department of Civil, Construction, \\ and Environmental Engineering\\
  University of Memphis\\
  Memphis, TN 38152  \\
  \texttt{cimeier@memphis.edu} \\
}

\begin{document}
\maketitle
\begin{abstract}
Since its publication in 1949, Langbein’s formula has been applied ubiquitously in both research documents and national guidelines concerning frequency analyses (FAs) of hydrologic extremes. Given a time series of independent peak-over-threshold (POT) events and the corresponding annual maxima (AM) series—defined as the subset of extremes representing the largest event in each year—the formula provides a theoretical relationship between the return period $T$ derived from the AM series and the average recurrence interval $ARI$ from the POT series, for any fixed event magnitude.
Despite the minimal assumptions required—specifically, that exceedance counts follow a homogeneous Poisson process—there are real-world situations where the validity of the formula may be compromised. Typical cases include non-stationary processes, mixed-event populations, and over- or under-dispersion in exceedance counts. In this work, we extend Langbein’s formula to account for these three cases. We demonstrate that, with appropriate adaptations to the definitions of $T$ and $ARI$, the traditional functional form of Langbein’s relationship remains valid for non-stationary processes and mixed populations. However, accounting for dispersion effects in exceedance counts requires a generalization of Langbein’s relationship, of which the traditional version represents a limiting case. 
\end{abstract}


\section{Introduction}
There are two established approaches for performing frequency analyses (FA) of extreme events: one based on counting the average number of years between successive annual maxima—known as the annual-maxima (AM) method—and another based on measuring the expected interarrival time between exceedances of a predefined threshold—referred to as the partial duration (PD) or peaks-over-threshold (POT) method~\cite[]{Coles2001}. More specifically, for a fixed magnitude $Q$, the AM method estimates the average number of years between annual maxima whose magnitude is equal to or greater than $Q$, referred to as the return period $T$ of the event. $T$ is defined as the expected number of uneventful years (i.e., with annual maxima below $Q$) before a "success"—a year in which the annual maximum is equal to or greater than $Q$—based on a homogeneous geometric distribution~\cite[i.e., with constant probability of success;][]{Salas2014}. 

In contrast, the PD method estimates the average recurrence interval—or $ARI$—of the event $Q$, defined as the expected interarrival time between two independent subsequent exceedances of $Q$, according to an excess count model—typically a Poisson process~\cite[]{Pan2022}—describing the annual number of threshold crossings. For example, it is well known that, if the number of exceedances follows a Poisson process, then the interarrival time between them is exponentially distributed. Appendix A provides a brief refresher on concepts related to traditional frequency analyses of extremes.

The AM and PD methods generally yield equivalent frequency estimates for events with recurrence intervals above about 10 years; in such cases, the theoretical difference between $T$ and $ARI$ approaches 0.5 years asymptotically~\cite[]{WangHolmes2020}, which is negligible for most practical purposes. However, in the range of more frequent events—say, with recurrence intervals between approximately 1 and 5 years—the two estimates diverge noticeably \cite[]{Langbein1949,Dellaira2025LangComm}. In this range, $T$ tends to overestimate the interarrival time between events, thereby underestimating their actual frequency of occurrence. This discrepancy arises because counting the number of years between annual maxima above $Q$ inevitably overlooks other occurrences within the same year; counting the number of years also prevents $T$ from capturing occurrences at the sub-annual scale~\cite[]{Dellaira2025LangComm}. 

Due to this limitation, much research and official guidelines from various countries—including Australia's handbook for peak flow estimation \cite[]{Ball2019} and the U.S. Bulletin 17C for flood frequency analysis \cite[]{England2018}—recommend using the PD method when predicting events with recurrence intervals below 10 years~\cite[]{Dellaira2025LangComm,PanRahman2022}. Inaccurate predictions of low-magnitude, high-frequency events can bias results in several disciplines, including river geomorphology, channel restoration, and stream ecology~\cite[]{Armstrong2012,Dellaira2025LangComm}.
Despite its advantages, the PD method requires continuous data and involves more complex preprocessing steps (e.g., threshold selection and the implementation of independence criteria), making it less accessible for practitioners. As a result, AM frequency analyses remain the preferred method in many applications~\cite[]{PanRahman2022,England2018}. A widely used workaround is to convert AM-based estimates into PD-based ones using the well-known relationship between $T$ and $ARI$—first derived by \cite{Langbein1949} (Equation~\ref{EqLang})—as proposed in, e.g., \cite{Dalrymple1960,Page1981,Beran1977,Bezak2014} and \cite{Olafsdottir2021}, among others. This correction method based on Langbein's relationship is also systematically implemented in the U.S. Rainfall Atlas 14~\cite[]{Perica2011}.

\begin{equation}
\frac{1}{T} = 1 - \exp\left(-\frac{1}{ARI}\right)
\label{EqLang}
\end{equation}

Langbein's formula (Equation~\ref{EqLang}) is strictly valid when the number of threshold exceedances in the PD series follows a homogeneous Poisson process~\cite[also see Appendix A]{Takeuchi1984}—i.e., with a constant rate of exceedances—a condition that holds in a wide range of applications~\cite[]{Coles2001}. Notably, no assumptions are made regarding the distribution of the magnitudes of AM and PD events, which enhances the general applicability of the formula, regardless of the specific characteristics of the AM and PD series~\cite[]{Dellaira2025LangComm}. Nevertheless, there are common circumstances where the assumption of a homogeneous Poisson process may be too restrictive, thereby undermining the validity of Equation~\ref{EqLang}.

For example, \cite{Armstrong2012} documented positive trends in the number of moderate floods in New England, which may violate the assumption of a constant exceedance rate. If the mean number of events presents temporal trends, a non-homogeneous Poisson process (NHPP)—i.e., one with a time-varying rate—may be more appropriate for modeling such series.

Another potential limitation of the formula’s theoretical foundation is that exceedances may result from multiple concurrent mechanisms, challenging the use of a single count model. An increasing body of research highlights the issue of mixed populations across various physical processes—including precipitation, flooding, and sea waves~\cite[]{Durn-Rosal2022,Barth2019,Villarini2016}—, suggesting that the assumptions that traditional frequency analysis methods rely on—including Poisson as a count model for the exceedances \cite[]{Coles2001}— are unrealistic and thus ill-suited for reliable predictions in real-world applications.

A third scenario where Langbein’s formula may not hold is in the presence of over- or under-dispersion in the number of exceedances—i.e., when the variance significantly exceeds or falls below the mean~\cite[]{Bezak2014}, thus violating the Poisson process assumption. A well-known example of over-dispersion is found in intermittent flows in arid regions, where alternating wet and dry years cause events to cluster~\cite[]{Zaman2012,Metzger2020}.

In summary, although scenarios in which Langbein’s formula may not be rigorously valid are evidently quite common—and deviations from the theoretical behavior prescribed by Equation~\ref{EqLang} are well documented in real-world applications \cite[see, e.g.,][]{Beran1977,Keast2013,Page1981,Takeuchi1984,Anderson1970}—no substantial efforts have yet been undertaken to systematically address these limitations. Non-stationarity, mixed populations, and dispersion in exceedance counts are among the most typical conditions that challenge the validity of Langbein’s formula (Equation \ref{EqLang}), by potentially rendering its core assumption too restrictive. These conditions can arise in applications concerned with frequent events, where the formula is most commonly used \cite[]{Dellaira2025LangComm,Perica2011,Williams1978}, including, e.g., frequency analyses of low-magnitude, high-frequency floods in geomorphology, channel restoration projects, economic risk assessments related to nuisance flooding, and studies of moderate but recurrent storms affecting coastal sediment budgets \cite[]{Armstrong2012,Lee1998ShorelineEvol,Moftakhari2017,Moftakhari2018}.

To address these limitations and enable modelers to reliably bridge $T$ and $ARI$ estimates from AM and PD frequency analyses even when the original formulation of Langbein's relationship (Equation~\ref{EqLang}) may not be valid, this work derives extensions of Langbein’s formula for three more general cases: non-stationarity (Section 2), mixed populations (Section 3), and dispersion in the number of exceedances (Section 4). Section~5 concludes with a discussion of the implications of these extensions.

\section{Extension to non-stationary processes}

Non-stationarity is observed when the drivers of extreme events vary on timescales comparable to the length of the observational record, causing the statistical properties of a process to change over time. To account for these effects, time- or covariate-dependent distribution parameters are introduced, allowing modelers to capture trends in the annual maxima (AM) and peaks-over-threshold (PD) series \cite[]{Coles2001}.

Contrary to the stationary case, where the return period $T$ is defined as the expected number of trials in a homogeneous geometric distribution with constant probability $p$ (see Appendix A), for a non-stationary process, the annual probability of exceedances $p$ varies with time, meaning that $T$ must be determined as the expected value from a non-homogeneous geometric distribution (see Appendix B for more details). By denoting this time-varying probability as $p_x$, representing the probability of observing an annual maximum above a reference magnitude $Q$ in year $x$, the return period $T_{NS}$ of $Q$ in a non-stationary context is given by Equation \ref{T_NS} \cite[also see Appendix B]{Salas2014}:

\begin{equation}
T_{NS}(t_0)=\sum_{x=t_0+1}^{+\infty} x\,p_x\prod_{k=t_0+1}^{x-1} (1-p_k)
\label{T_NS}
\end{equation}

Because the probability $p_x$ of an annual maximum exceeding $Q$ varies from year to year, $T_{NS}$ depends on the starting year $t_0$ of the count.

Similarly, $ARI_{NS}$ under non-stationary conditions can be viewed as a generalization of the stationary case. Specifically, in stationary processes, $ARI$ is the expected waiting time derived from a homogeneous Poisson process—i.e, with a constant rate of exceedances. Under non-stationary conditions, however, the exceedance rate $\lambda(t)$ is a function of time. As a result, $ARI_{NS}$ represents the expected waiting time associated with a non-homogeneous Poisson process—i.e., with a time-varying exceedance rate—, as illustrated in Equation \ref{ARI_NS} (see derivations in Appendix B):

\begin{equation}
\begin{split}
ARI_{NS}(t_0)=\int_{0}^{\infty} w\, \lambda(t_0+w)\,  \exp{\left[ -\int_{0}^{w}\lambda\left(t_0+s\right)\, ds \right]}\, dw
\label{ARI_NS}
\end{split}
\end{equation}

Here, $w$ denotes realizations of the waiting time until the first occurrence after $t_0$, measured from $t_0$. As such, $w$ is non-negative. If we indicate the time of the first occurrence as $t_1$, then $t_1 = t_0 + w$. Like $T_{NS}$, $ARI_{NS}$ is a function of the initial time $t_0$. An important difference between the two methods is that $T_{NS}$ is calculated for a discrete timeline, using annual blocks ($x=t_0+1,\, t_0+2, ...$), while $ARI_{NS}$ considers continuous time. To unify the two frameworks, here we calculate the annual probability of exceedance $p_x$ by integrating the instantaneous excess rate over each year; this is equivalent to considering yearly average annual-exceedance probabilities. Additionally, we measure $ARI$ starting from the beginning of the initial year $t_0$ (see Appendix B, Equations \ref{ARI_NS1} and \ref{T_NS1_stepInt} or Equations \ref{ARI_NS2} and \ref{T_NS2_stepInt}).

\begin{figure}[htbp!] 
\noindent
\includegraphics[width=0.9\textwidth]{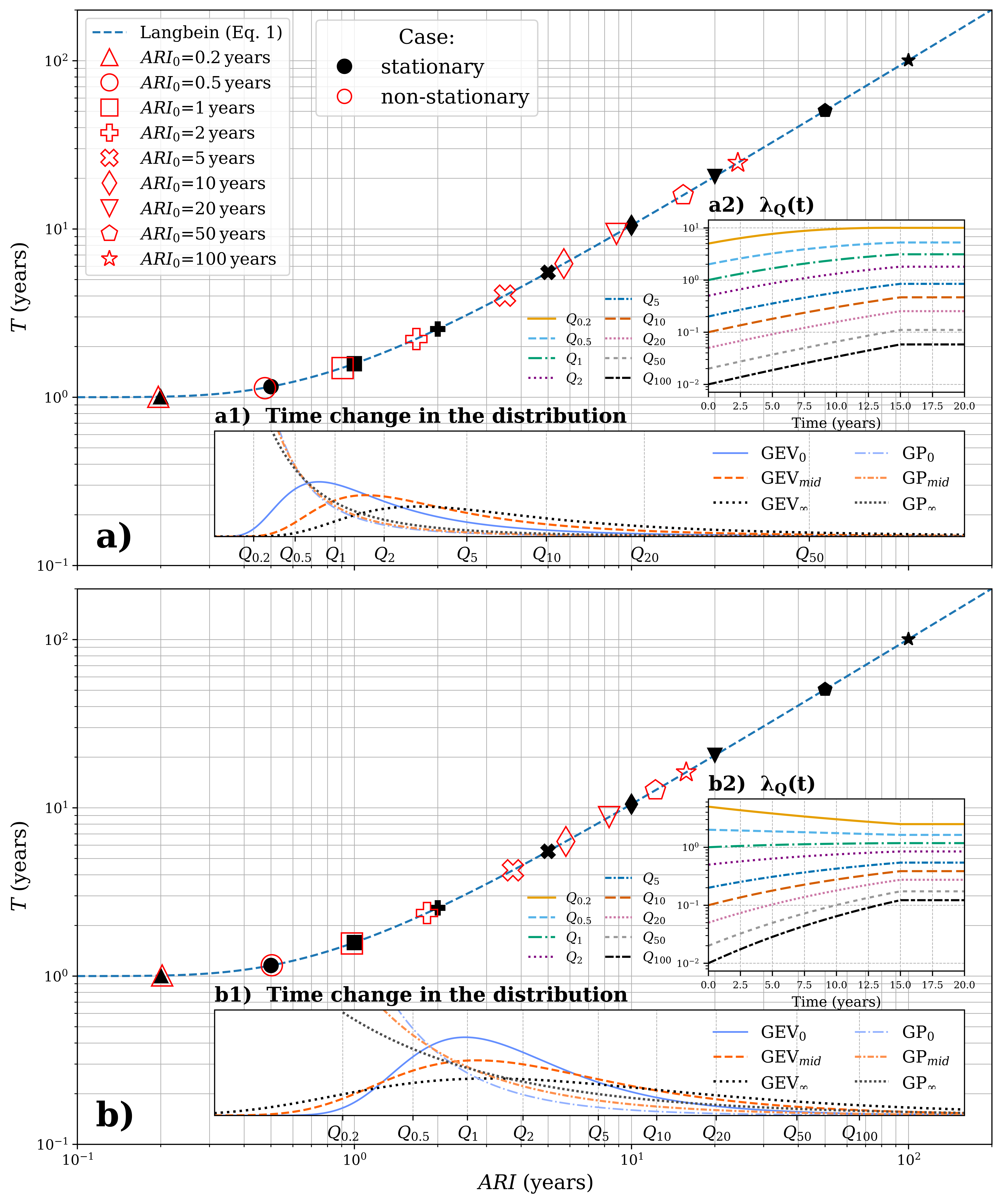}
\caption{For two distinct examples of non-stationary processes, a) and b), comparison of $T_{NS}(t_0)$-$ARI_{NS}(t_0)$ pairs, estimated at year $t_0$ under non-stationary conditions, with the corresponding $T$-$ARI$ pairs if conditions at year $t_0$ were stationary. Fixed-magnitude events $Q_{0.2}$, $Q_{0.5}$, $Q_{1}$, $Q_{2}$, $Q_{5}$, $Q_{10}$, $Q_{20}$, $Q_{50}$, $Q_{100}$ have an $ARI_0$—the average recurrence interval at time $t_0$ if the process were stationary—of 0.2, 0.5, 1, 2, 5, 10, 20, 50, and 100 years, respectively. Panels a1) and b1) outline the assumed shift in the distribution of AM and PD series due to non-stationarity, by plotting the curves at time $t_0$ (GEV$_0$ and GP$_0$), $t_0$+7.5 years (GEV$_{mid}$ and GP$_{mid}$), and $t_0$+15 years (GEV$_\infty$ and GP$_\infty$). After $t_0$+15 years, it is assumed that the distributions remain stationary in both examples. Panels a2 and b2 show the resulting temporal changes in the mean rate of exceedances, $\lambda_Q(t)$, of the fixed-magnitude events $Q_{0.2}$, $Q_{0.5}$, $Q_{1}$, $Q_{2}$, $Q_{5}$, $Q_{10}$, $Q_{20}$, $Q_{50}$, and $Q_{100}$, respectively. Panels a1) and b1) mark the event magnitudes along the x-axis. More information about these two examples are provided in Appendix C.}
\label{FigNS}
\end{figure}

To illustrate the effects of non-stationarity on the relationship between $T_{NS}$ and $ARI_{NS}$, we consider two arbitrary yet realistic examples of non-stationary processes with time-varying distribution parameters (panels a1 and b1 in Figure \ref{FigNS}). We estimate $T_{NS}$ and $ARI_{NS}$ for a range of fixed-magnitude events with values of $ARI_0$—the average recurrence interval at time $t_0$ if the process were stationary—of 0.2, 0.5, 1, 2, 5, 10, 20, 50, and 100 years, and compare them with their counterparts obtained under stationary conditions. Following established modeling approaches for non-stationary processes \cite[]{Olafsdottir2021,Coles2001,katz2012statistical}, both examples in Figure \ref{FigNS} assume a linear temporal variation in the location parameter and an exponential variation in the scale parameter of the distribution of annual maxima, using rates of change consistent with real-world observations (see Appendix C for more details).

In the first example (Figure \ref{FigNS}a), the distribution of AM experiences a shift to the right, induced by a strong positive trend in its location parameter, accompanied by a relatively less marked increase in the spread of the distribution and consequent lowering of its peak, reflecting a concurrent growth in its scale parameter over time. As a result, the rate of exceedances $\lambda_Q$ increases systematically for all fixed-magnitude events in Figure \ref{FigNS}a, until a plateau is reached after about 15 years, as illustrated in panel a2. On the other hand, in the second example (Figure \ref{FigNS}b) the increase in the spread of the distribution of AM is more pronounced than the shift of its location (panel b1), together resulting in a decrease in the rate of occurrence of small-recurrence-interval events and an increase in the rate of occurrence of the more extreme ones (panel b2). The distribution of the PD series in the two examples changes consistently with the temporal variations in the corresponding distribution of AM and the assumed change in the rate of threshold exceedances, following the duality between the two types of distributions \cite[also see Appendix C]{WangHolmes2020}. For the sake of illustrating the effects of non-stationarity even for the smaller quantiles, in these two examples we have exacerbated the variability in the rate of threshold exceedances, by introducing an increase up to as twice as the original rate (Figure \ref{FigNS}a) and a decrease up to half the original rate of exceedances (Figure \ref{FigNS}b), respectively, within a 15-year period. The initial excess rate was assumed equal to five times a year for both examples.

There are two key takeaways from our examples in Figure \ref{FigNS}. First, the $T_{NS}$-$ARI_{NS}$ pairs, all estimated at the same time $t_0$ for a range of fixed-magnitude events with given (stationary) $ARI_0$ at year $t_0$, all lie on Langbein's theoretical curve defined by Equation \ref{EqLang}, even though they do not occupy the same position along the curve as they would under stationary conditions (compare black empty markers with red empty ones in Figure \ref{FigNS}). This, in turn, suggests that Langbein's equation remains valid under non-stationary conditions, provided that $T_{NS}(t_0)$ and $ARI_{NS}(t_0)$ are computed by accounting for the non-stationary nature of the process—i.e., using Equations \ref{T_NS} and \ref{ARI_NS}. The second takeaway from our example concerns how the $T_{NS}$-$ARI_{NS}$ pairs (red empty markers in Figure \ref{FigNS}) are shifted relative to their stationary $T$-$ARI$ counterparts (black filled markers). Depending on whether the rate of occurrences $\lambda_Q(t)$ of the generic event $Q$ increases or decreases starting from $t_0$, $T_{NS}(t_0)$ and $ARI_{NS}(t_0)$ are either anticipated or delayed compared to the corresponding $T$ and $ARI$ under stationary conditions, as indicated by a shift to the left or right, respectively. Greater rates of change lead to stronger deviations from the stationary $T$-$ARI$ points; moreover, the effects of non-stationarity are inevitably more pronounced for larger $T$-$ARI$ values due to the longer time scales over which those effects accumulate.

To wrap up our discussion on the effects of non-stationarity on Langbein's formula, we can conclude that, assuming that threshold exceedances follow a non-homogeneous Poisson process with time-varying parameter $\lambda(t)$—this is a generalization of the assumption of a Poisson process with constant annual rate of exceedances—Langbein's formula is still valid for non-stationary processes and can be re-written as shown in Equation \ref{EqNSLang}:

\begin{equation}
\frac{1}{T_{NS}(t_0)} = 1 - \exp\left(-\frac{1}{ARI_{NS}(t_0)}\right)
\label{EqNSLang}
\end{equation}

The difference in notation between Equations \ref{EqLang} and \ref{EqNSLang} highlights that return period and average recurrence interval are derived under different formulations—for stationary and non-stationary processes, respectively—and that, in the latter case, they depend on the year $t_0$ at which they are estimated. By calculating $T_{NS}$–$ARI_{NS}$ pairs for a range of fixed-magnitude quantiles and plotting them on the $ARI$-$T$ plane, we obtained empirical evidence supporting the validity of Equation \ref{EqNSLang}—i.e., the observation that $T_{NS}$–$ARI_{NS}$ pairs still lie on the theoretical curve provided by Equation \ref{EqLang}. It is also possible to analytically demonstrate the validity of Langbein's formula under non-stationarity (Equation \ref{EqNSLang}), regardless of any specific temporal variations in the distribution parameters, as shown in Appendix D.

\section{Extension to mixed populations}

Mixed populations refer to situations where the observed data originate from more than one underlying process or distribution. A typical example is a series of peak flows at a given river section, where some peaks—clustered in a relatively short period after the onset of the warm season—may be predominantly generated by snowmelt and-rain-on-snow events, while other peaks occurring at different times of the year may be attributable to precipitation only \cite[]{Berghuijs2016}. Among the latter, it may be possible to further distinguish event types based on the atmospheric drivers of precipitation—e.g., frontal systems, convective cells, atmospheric rivers, or tropical cyclones, depending on local climate \cite[]{Barth2017,Villarini2016}—each with major differences in terms of rainfall amounts, durations, frequency, and spatial scale.

Another typical example is provided by sea wave heights, where individual events may result from distinct physical mechanisms such as locally generated wind waves, remotely generated swell, wave interactions in mixed sea states, or secondary waves caused by interactions with obstacles. These mechanisms differ significantly in terms of energy source, spatial coherence, and temporal persistence, often leading to multimodal distributions in wave height data \cite[]{huang2020joint,Durn-Rosal2022}.

As recalled in Appendix A, Langbein's formula is obtained under the assumption that all exceedances follow a Poisson process with a constant parameter $\lambda$. Under mixed populations, this assumption may not hold, as different mechanisms may generate events at different rates, regardless of any differences in the distributions of magnitudes of those diverse event types. To address this limitation, we assume that events from each distinct generating mechanism can be modeled as a separate Poisson process with parameter $\lambda_i$.

A fundamental property of Poisson processes is that the superposition (i.e., sum) of $n$ independent Poisson processes, each with rate $\lambda_i$, results in another Poisson process with a rate equal to the sum of the individual rates \cite[]{kingman1992poisson}:

\begin{equation}
\begin{split}
\lambda_{MP} = \sum_{i=1}^n \lambda_i
\label{HPPsuperposition}
\end{split}
\end{equation}

In the above Equation \ref{HPPsuperposition}, we introduced $\lambda_{MP}$, which denotes the parameter of the Poisson process describing the number of exceedances arising from mixed generating mechanisms. This property holds under the assumption that the individual processes are independent and that events from different processes are distinguishable \cite[i.e., the same event does not result from two co-existing mechanisms;][]{kingman1992poisson}. Because the number of mixed exceedances still follows a Poisson process with constant parameter $\lambda_{MP}$, the formulations illustrated in Appendix A regarding $T$, $ARI$, and their relationship can all be extended straightforwardly to the mixed populations case, hence justifying the following Equation \ref{EqMPLang}. 

\begin{equation} 
\frac{1}{T_{MP}} = 1 - \exp\left(-\frac{1}{ARI_{MP}}\right) 
\label{EqMPLang} 
\end{equation} 

What truly changes with respect to the traditional formula (Equation \ref{EqLang}) is that $T_{MP}$ and $ARI_{MP}$ should not be estimated using a traditional, unimodal statistical distribution, as such a model would not be able to capture the intrinsic multivariate nature of the process. To clarify this, let us refer to the method outlined in \cite{WangHolmes2020}. Assuming that the events in our dataset can arise from two distinct generating mechanisms, each following a Poisson process with annual rates of exceedance $\lambda_1$ and $\lambda_2$, respectively, we can estimate the annual probabilities $P(Q_1 \le q) = \exp(-\lambda_1)$ and $P(Q_2 \le q) = \exp(-\lambda_2)$ for the two types of events—denoted as $Q_1$ and $Q_2$—from the two independent processes, and combine them as follows:

\begin{equation}
\begin{split}
P(Q \le q) = P(Q_1 \le q) \, P(Q_2 \le q)
\label{combinedNonExcProb}
\end{split}
\end{equation}

In the above Equation \ref{combinedNonExcProb}, $P(Q \le q)$ denotes the probability that, in a given year, the annual maximum $Q$ does not surpass the reference value $q$, irrespective of the generating mechanism of that event. By switching to probabilities of exceedance, we obtain:

\begin{equation}
\begin{split}
P(Q > q) = 1-(1-P(Q_1 > q)) \, (1-P(Q_2 > q))= \\ =
P(Q_1 > q)+P(Q_2 > q)-P(Q_1 > q) \, P(Q_2 > q)
\label{combinedExcProb}
\end{split}
\end{equation}

Finally, from Equation \ref{combinedExcProb} and the definition of return period (Equation \ref{EqT}), we can derive the following expression for the return period $T_{MP} = 1 / P(Q > q)$ of an annual maximum exceeding $q$, regardless of its generating mechanism, as a function of the return periods of events from the two generating processes considered separately, as shown in Equation \ref{EqT_MP} \cite[]{WangHolmes2020}:

\begin{equation}
\begin{split}
T_{MP} = \left( \frac{1}{T_1} + \frac{1}{T_2} - \frac{1}{T_1} \, \frac{1}{T_2} \right)^{-1}
\label{EqT_MP}
\end{split}
\end{equation}

By individually applying Langbein's formula to express $T_1$ as a function of $ARI_1$, $T_2$ as a function of $ARI_2$, and $T_{MP}$ as a function of $ARI_{MP}$, we can rewrite the two sides of Equation \ref{EqT_MP} as follows:

\begin{equation}
\begin{split}
\frac{1}{T_{MP}} = 1 - \exp \left( -\frac{1}{ARI_{MP}} \right)
\label{leftside}
\end{split}
\end{equation}

\begin{equation}
\begin{split}
\left[1 - \exp \left( -\frac{1}{ARI_1} \right) \right] + \left[1 - \exp \left( -\frac{1}{ARI_2} \right) \right] - \left[1 - \exp \left( -\frac{1}{ARI_1} \right) \right] \left[1 - \exp \left( -\frac{1}{ARI_2} \right) \right] = \\
= 1 - \exp \left( -\frac{1}{ARI_1} \right) \exp \left( -\frac{1}{ARI_2} \right) = \\
= 1 - \exp \left( - \left[ \frac{1}{ARI_1} + \frac{1}{ARI_2} \right] \right)
\label{rightside}
\end{split}
\end{equation}

By substituting Equations \ref{leftside} and \ref{rightside} into Equation \ref{EqT_MP}, we obtain the expression for $ARI_{MP}$ given in Equation \ref{EqARI_MP} \cite[]{WangHolmes2020}:

\begin{equation}
\begin{split}
\frac{1}{ARI_{MP}} = \frac{1}{ARI_1} + \frac{1}{ARI_2}
\label{EqARI_MP}
\end{split}
\end{equation}

This result is in line with our expectations: considering that $ARI_1 = \lambda_1^{-1}$ and $ARI_2 = \lambda_2^{-1}$, Equation \ref{EqARI_MP} shows that $ARI_{MP}$ is equal to the inverse of $\lambda_{MP} = \lambda_1 + \lambda_2$, consistent with the fact that it represents the expected waiting time associated with a Poisson process with mean rate $\lambda_{MP} = \lambda_1 + \lambda_2$ \cite[from the superposition property of Poisson processes;][]{kingman1992poisson}). It can be immediately derived that $ARI_{MP} < \min\left(ARI_1, \ ARI_2 \right)$ \cite[]{WangHolmes2020}, consistent with the fact that $ARI_{MP}$ is the expected waiting time for the mixed process, while $ARI_1$ and $ARI_2$ describe recurrence intervals within each individual process.

In summary, as $T_{MP}$ and $ARI_{MP}$ are generally different from estimates of $T$ and $ARI$ obtained from traditional frequency analysis methods (i.e., that ignore mixed populations), the risk for bias under mixed populations is mainly related to not using suitable statistical models that can capture multi-modality, rather than to the validity of Langbein's formula (Equation \ref{EqLang}). Obviously, using Equation \ref{EqLang} to convert a biased estimate of $T$ into an estimate of $ARI$ will only propagate the error, but the issue lies not in the validity of the formula itself. Provided that $T_{MP}$ (or $ARI_{MP}$) is correctly estimated by explicitly accounting for mixed populations, Langbein's relationship to obtain the corresponding $ARI_{MP}$ (or $T_{MP}$) remains valid, under the assumption that the number of exceedances from each individual mechanism follows a Poisson process—or, equivalently, that the mixed process itself follows a Poisson process with cumulative rate $\lambda_{MP}$. The version given in Equation \ref{EqMPLang} is formally equivalent to Equation \ref{EqLang} and the change in notation only serves to emphasize the importance of correctly adapting the statistical models used to estimate $T_{MP}$ and $ARI_{MP}$ to reflect the presence of mixed populations.

We conclude this section with an important clarification regarding the expression for $T_{MP}$ given in Equation \ref{EqT_MP}. The use of that equation assumes that events from the two distinct generating mechanisms are preliminarily separated, and that two distinct frequency analyses are performed independently for each group of events using standard (i.e., unimodal) statistical models, yielding $T_1$ and $T_2$ (for the same fixed reference magnitude $q$). However, if one uses statistical models that are able to account for mixed populations from the outset, such as weighted mixed distributions \cite[]{Barth2019,Durn-Rosal2022}, $T_{MP}$ can be estimated directly as the inverse of the probability of exceedance provided by that model (Equation \ref{EqT}), and Langbein's formula can then be used to obtain the corresponding $ARI_{MP}$.

\section{Extension to processes exhibiting over- or under-dispersion}

The cornerstone assumption behind Langbein's formula is that the number of exceedances follows a Poisson process \cite[]{WangHolmes2020,Takeuchi1984}. Notably, the two extensions discussed in Sections 2 and 3—addressing non-stationarity and mixed populations—represent generalizations of this assumption: namely, a non-homogeneous Poisson process with a time-varying rate of exceedances, and a superposition of distinct Poisson processes associated with different coexisting generating mechanisms, both of which still constitute Poisson processes. In this section, we will explore alternative count models that are better suited to cases of over- or under-dispersion in the number of exceedances, and examine their effects on the relationship between $T$ and $ARI$.

Two alternatives to the Poisson count model are the Binomial and Negative Binomial distributions (Equations \ref{Bin} and \ref{NegBin}, respectively), suitable in cases where the variance $V$ of the annual number of exceedances is smaller (under-dispersion) or larger (over-dispersion) than the mean $\lambda$, respectively \cite[]{Bezak2014,Bhunya2012,Cunnane1979,Onoz2001}. By considering these alternative count models, the variance $V$ of the number of exceedances can therefore vary with respect to  $\lambda$, and the difference between $V$ and $\lambda$ is controlled by either the parameter $r$ in the Negative Binomial model (over-dispersion; see Appendix E) or the parameter $\gamma$ in the Binomial model (under-dispersion; see Appendix F) as illustrated in Equation \ref{EqVar}: 

%
%


\begin{equation}
\begin{split}
V =
\left\{
\begin{array}{ll}
\displaystyle \lambda + \frac{\lambda^2}{r} & \text{over-dispersion} \\\\
\displaystyle \lambda - \frac{\lambda^2}{\gamma} & \text{under-dispersion}
\end{array}
\right.
\label{EqVar}
\end{split}
\end{equation}

Both $r$ and $\gamma$ can be readily obtained from the PD series as functions of the mean number $\lambda$ and variance $V$ of the number of threshold exceedances—specifically, both are given by the ratio between $\lambda^2$ and $|V-\lambda|$ (Equations \ref{NegBinR} and \ref{BinG}) and are strictly positive. $V$ tends towards $\lambda$ (i.e, equi-dispersion) as $r\rightarrow+\infty$ or $\gamma\rightarrow+\infty$ \cite[]{Zhou2015}. It is worth noting here that Equation \ref{EqVar} is valid for any $\lambda$—i.e., for any high threshold associated with that $\lambda$, as illustrated in Appendix G. In the second part of Equation \ref{EqVar}, $\gamma$ must be greater than $\lambda$ to ensure $V$ is positive; this condition is always met, since $\gamma>\lambda_u$—where $\lambda_u$ is the rate of exceedances associated with the original threshold $u$ used to define the PD series—and for any magnitude $Q>u$ with an exceedance rate of $\lambda$ the condition $\lambda_u>\lambda$ always holds.

Expressions for deriving the return period $T$ for processes exhibiting over-dispersion and under-dispersion are derived in Appendices E and F, respectively (Equations~\ref{EqT_NB} and~\ref{EqT_B}). By combining these results, we obtain a piece-wise formulation for $T$, presented in Equation~\ref{EqT_Gen}:
%
%

\begin{equation}
\begin{split}
T =
\left\{
\begin{array}{ll}
\displaystyle \frac{1}{1-(\lambda/V)^r} = \frac{1}{1-(1+\lambda/r)^{-r}} & \text{over-dispersion} \\\\
\displaystyle \frac{1}{1-(\lambda/V)^{-\gamma}} = \frac{1}{1-(1-\lambda/\gamma)^{\gamma}} & \text{under-dispersion}
\end{array}
\right.
\label{EqT_Gen}
\end{split}
\end{equation}

The last terms of both expressions in Equation \ref{EqT_Gen} are obtained by substituting Equation \ref{EqVar}. Notably, the return period derived for homogeneous Poisson processes (Equation~\ref{EqT}) emerges as a limiting case of Equation~\ref{EqT_Gen} as $V\rightarrow\lambda$ (or, equivalently, as $r\rightarrow0$ or $\gamma\rightarrow0$), as detailed in Appendices E and F. 

We now introduce a unified dispersion parameter $\psi$, defined in Equation \ref{EqPsi}, to  control for both over- or under-dispersion in the number of exceedances:

%
%

\begin{equation} 
\begin{split} 
\psi = 
\left\{
\begin{array}{ll}
\displaystyle \frac{1}{r} & \text{over-dispersion} \\
\displaystyle -\frac{1}{\gamma} & \text{under-dispersion}
\end{array}
\right.
\label{EqPsi} 
\end{split} 
\end{equation} 

By substituting Equation \ref{EqPsi} into Equation \ref{EqVar}, we derive a general relationship between $\lambda$ and $V$, as presented in Equation \ref{Var_lambda_psi}, which holds for processes exhibiting any degree of dispersion in exceedance counts. From this relationship, it becomes clear that positive increments in $\psi$ amplify over-dispersion, whereas negative shifts promote under-dispersion. Equi-dispersion is a special case corresponding to $\psi=0$.

\begin{equation}
V = \lambda + \psi \lambda^2
\label{Var_lambda_psi}
\end{equation}

We can also reparametrize Equation~\ref{EqT_Gen} in terms of $\psi$, yielding the general formulation for $T$ shown in Equation~\ref{EqT_Gen_psi}. The return period for equi-dispersed processes (Equation~\ref{EqT}) then emerges as a limiting case of this formulation as $\psi \rightarrow 0$.

\begin{equation}
\begin{split}
\displaystyle T= \frac{\displaystyle 1}{\displaystyle 1- \left(1+\displaystyle \psi \lambda \right)^{-1/\psi}}
\label{EqT_Gen_psi}
\end{split}
\end{equation}

We are now ready to extend Langbein’s formula to processes exhibiting over- or under-dispersion. Regardless of the count model, as long as the exceedance rate remains constant, the expected interarrival time between two subsequent exceedances—i.e., the $ARI$—is simply the inverse of the mean rate of exceedances, $\lambda$. By substituting $ARI$ into Equation~\ref{EqT_Gen_psi}, we obtain the following generalization of Langbein’s formula, presented in Equation~\ref{EqGenLang}:

\begin{equation}
\begin{split}
\displaystyle T= \frac{\displaystyle 1}{\displaystyle 1- \left(1+\frac{\displaystyle \psi}{ \displaystyle ARI} \right)^{-1/\psi}}
\label{EqGenLang}
\end{split}
\end{equation}

\begin{figure}[htbp!] 
\noindent
\includegraphics[width=\textwidth]{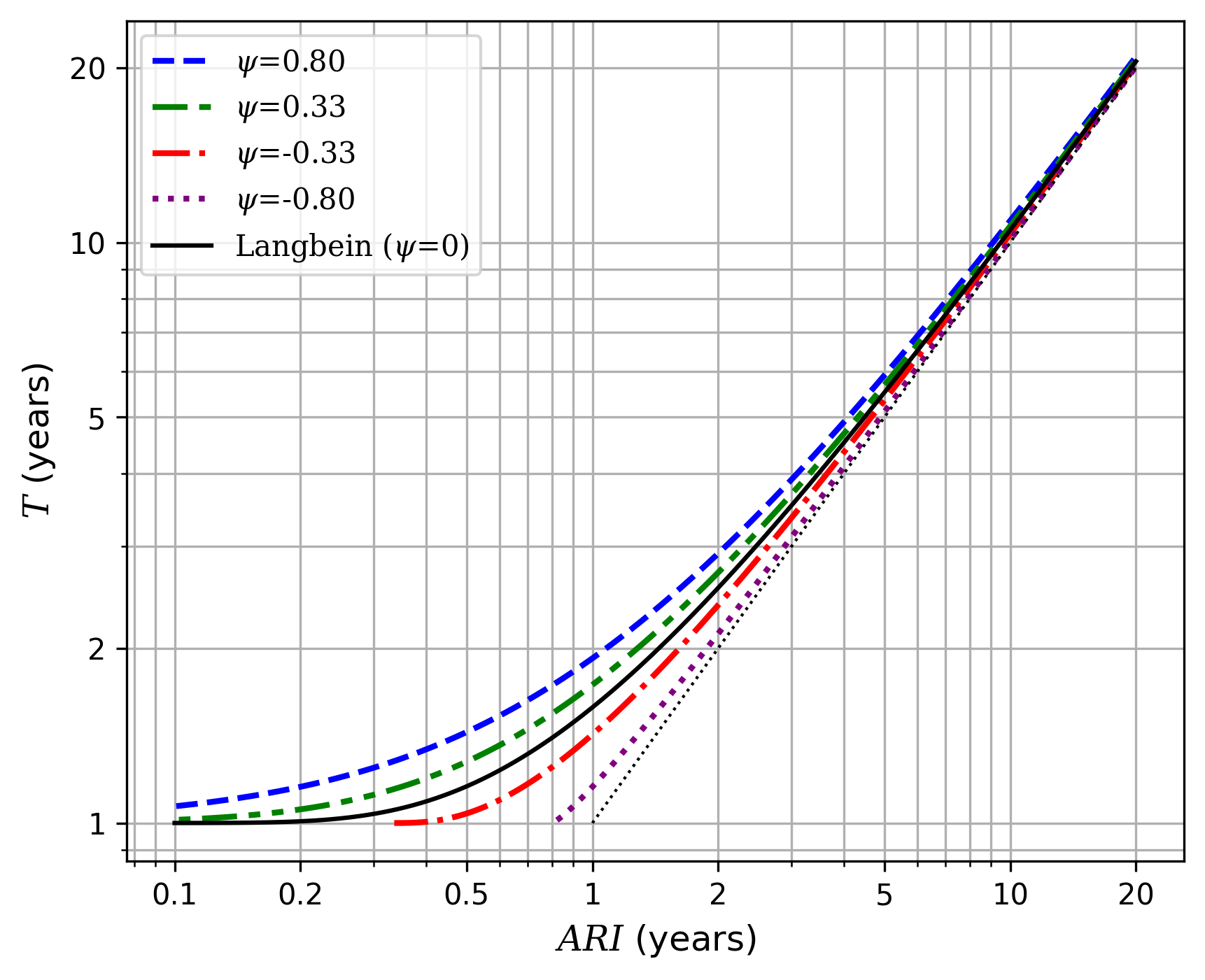}
\caption{Change in the relationship between $T$ and $ARI$ under over- and under-dispersion in the exceedance count. A gradually more positive dispersion parameter $\psi$ is associated with increasing over-dispersion, while a gradually more negative $\psi$ indicates stronger under-dispersion. The values of $\psi$ that are larger in absolute value (0.80 and -0.80) correspond to cases of marked over- or under-dispersion, where the variance in the rate of exceedances is 1.8 times larger ($\psi=0.8$) or 5 times smaller ($\psi=-0.8$) than the mean rate of exceedances, when the latter is equal to 1. On the other hand, the values of $\psi$ of 0.33 and -0.33 correspond to more moderate cases of over- and under-dispersion; considering again a unit mean rate of exceedances, the variance is 1.33 times higher or 1.5 times lower, respectively.}
\label{FigLangGen}
\end{figure}

Equation~\ref{EqGenLang} generalizes Langbein’s formula to account for over- and under-dispersion in the number of exceedances, modeled by positive and negative values of $\psi$, respectively. Equi-dispersion corresponds to the limiting case $\psi \rightarrow 0$, in which Equation~\ref{EqGenLang} reduces to Langbein’s original formulation (Equation~\ref{EqLang}). Figure~\ref{FigLangGen} illustrates how the relationship between $T$ and $ARI$ varies across four different levels of dispersion, each associated with a distinct value of $\psi$, and compares these curves to Langbein’s original relationship. A thin 45-degree reference line is also included for visual comparison. Over-dispersion causes an upward shift in the portion of the curve associated with frequent events, while under-dispersion results in a downward shift in that region. In contrast, all curves converge in the domain of rarer events, indicating that $T$ and $ARI$—and thus AM and PD frequency analyses—yield similar estimates for large, infrequent events, regardless of the degree of dispersion in the data.

Literature case studies have often reported deviation from the theoretical behavior outlined by Langbein's formula \cite[e.g.,][even including Langbein (1949) himself]{Beran1977,Page1981,Takeuchi1984,Keast2013}, with $T$-$ARI$ pairs obtained from concurrent AM and PD frequency analyses on the same datasets plotting above the curve in the range of frequent events—i.e., where the deviations between $T$ and $ARI$ are most pronounced. By considering the generalization proposed in Equation \ref{EqGenLang}, a possible interpretation of this reported behavior is over-dispersion in the original data.

Over-dispersion can be found in processes where events tend to cluster in certain years \cite[]{Bezak2014}. For example, highly over-dispersed flood exceedances are typically observed in arid and semi-arid regions \cite[]{Metzger2020}, which are known for greater variability in precipitation and flood occurrences than humid regions, due to the alternance of dry years with years with multiple events \cite[]{Zaman2012}. Interestingly, some of the case studies documenting consistent departures from Langbein's equation, such as \cite{Page1981} and \cite{Keast2013}, happen to involve arid watersheds. On the other hand, under-dispersion in exceedance counts can occur in processes where events are highly regularly spaced in time \cite[]{Bezak2014}, such as moderate floods in regulated rivers, due to controlled releases.

\FloatBarrier

\section{Final Remarks}
We derived three extensions of Langbein’s formula to account for non-stationarity, mixed populations, and over- or under-dispersion in exceedance counts. In the first two generalizations, we considered a non-homogeneous Poisson process and a superposition of distinct Poisson processes, respectively. Both approaches yield the same $T$–$ARI$ curve as the traditional relationship given in Equation~\ref{EqLang}. For a fixed event magnitude, the return period and average recurrence interval under non-stationarity or mixed populations—denoted as $T_{NS}$ and $ARI_{NS}$, or $T_{MP}$ and $ARI_{MP}$, respectively—will differ from the traditional estimates obtained under the assumption of a homogeneous Poisson process. However, the shifted $T_{NS}$–$ARI_{NS}$ (or $T_{MP}$–$ARI_{MP}$) pairs will still lie on Langbein’s curve, provided that the predictions are properly derived using models capable of capturing trends or multimodality in the data. In contrast, if traditional frequency analysis methods—relying on unimodal distributions and/or models with constant parameters—are improperly applied to non-stationary processes or mixed datasets, the resulting estimates of $T$ and $ARI$ may be biased and fail to conform to Equation~\ref{EqLang}, with stronger biases arising in the presence of particularly pronounced temporal trends or marked multimodality.

Dispersion in exceedance counts, on the other hand, does cause shifts in the curve describing the relationship between $T$ and $ARI$, particularly in the range of small-recurrence-interval events. The curve moves gradually upward for processes with increasing over-dispersion, and downward for those with increasing under-dispersion. Several studies have reported patterns in which empirical data points systematically lie above the theoretical curve. In light of the present findings, a plausible explanation for such observations is the presence of over-dispersion in the data.

In conclusion, we have demonstrated that Langbein’s formula remains relatively robust under non-stationarity and mixed populations, provided that (1) the exceedance rate can be modeled using a non-homogeneous Poisson process, or that the number of events from distinct generating mechanisms can be independently modeled using Poisson processes, and (2) the return period and average recurrence interval are properly estimated by accounting for trends and multimodality in the data. However, if the process exhibits strong dispersion—thus violating the assumption of Poisson-distributed exceedance counts—then modelers should consider using the generalized formulation proposed in this study (Equation \ref{EqGenLang}).

\appendix

\section{Return period, average recurrence interval, and the derivation of Langbein's formula}

\renewcommand{\theequation}{A\arabic{equation}}
\setcounter{equation}{0}

In this appendix, we will first recall the definitions of return period $T$ and average recurrence interval $ARI$, and then illustrate the derivation of Langbein's formula (Equation \ref{EqLang}) outlined in \cite{WangHolmes2020}. Consider a partial duration series of exceedances of a threshold $Q$ following a Poisson process with constant parameter $\lambda$ (Equation \ref{EqHPP})—also referred to as a homogeneous Poisson process (HPP).

\begin{equation}
P_{\text{Poisson}}(N(t)=k,\lambda) = \frac{e^{-\lambda\ t}[\lambda\ t]^k}{k!}
\label{EqHPP}
\end{equation}

Here, $N(t)$ denotes the number of threshold exceedances in the generic time interval [$0$-$t$). The expected number of exceedances in that interval is given by the product $\lambda\, t$; since $t$ is treated as a continuous variable expressed in years, the expected number of events in any given year is equal to $\lambda$.

The cumulative distribution function of the waiting time $W$ until the next exceedance is given in Equation \ref{cdfWaitingTimeHPP} \cite[]{Streit2010}, where $w$—a realization of $W$—is non-negative:

\begin{equation} 
\begin{split} 
F_{W}(w) = P(W \leq w) = 1 - P(W > w) = \\ = 1 - P(N(w) = 0) = \\ = 1 - \exp\left(-\lambda w\right)
\label{cdfWaitingTimeHPP} 
\end{split} 
\end{equation} 

In the above equation, we considered that the probability $P(N(t)=0)$ is equal to $\exp\left[-\lambda w\right]$ for a HPP. Differentiating $F_{W}(w)$ with respect to $w$, one can obtain the probability density function of waiting times, $f_{W}(w)$, given in Equation \ref{pdfWaitingTimeHPP}:

\begin{equation} 
f_W(w) = \lambda \exp\left(-\lambda w\right)
\label{pdfWaitingTimeHPP}
\end{equation}

The average recurrence interval $ARI$ is the expected value of $W$, and its expression is given in Equation \ref{EqARI1}:
 
\begin{equation} 
\begin{split} 
ARI = \mathbb{E}[W] =  \int_0^\infty w\, \lambda  \exp\left(-\lambda w\right)\, dw
\label{EqARI1} 
\end{split} 
\end{equation}

Integrating by parts as the integral of the product of $f(w)=w$ and $g'(w)=f_W(w)$, Equation \ref{EqARI1} can be re-written as shown in Equation \ref{EqARI2}:

\begin{equation} 
\begin{split} 
ARI = \int_0^\infty f(w)\ g'(w)\ dw = \\ 
\left[f(w)\ g(w)\right]_0^\infty-\int_0^\infty f'(w)\ g(w)\ dw = \\
=\left[-w(1-F_W(w) \right]_0^\infty+ \int_0^\infty \left[1-F_W(w) \right]dw = \\ = \int_0^\infty \exp{\left(-\lambda w\right)}dw = \\ =
\left[ -\frac{1}{\lambda}\,\exp{\left(-\lambda\ w \right)}\right]_0^\infty=\frac{1}{\lambda}
\label{EqARI2} 
\end{split} 
\end{equation} 

The boundary term $\left[-w(1-F_W(w) \right]_0^\infty$ in the third line of Equation \ref{EqARI2} vanishes for large $w$ because the exponential term $\exp{\left(-\lambda\ w\right)}$ from Equation \ref{cdfWaitingTimeHPP}—or, equivalently, the survival function $(1-F_W(w))$—decays faster than $w$ grows. In conclusion, Equation \ref{EqARI2} shows the well-known fact that the $ARI$ of a threshold exceedance with a constant annual rate of occurrence is equal to the inverse of that rate. Equation \ref{EqARI2} also illustrates that $ARI$ can alternatively be obtained as the integral of the survival function $S_W(w)$ of the waiting time $W$, defined as $S_W(w)=P(W>w)=1-F_W(w)$ \cite[]{Moore2016}.

Now we examine the definition of return period. The number $N_w$ of years before an annual maximum greater than $Q$ follows a geometric distribution with probability of success $p$=$1-\exp{(-\lambda)}$ \cite[]{WangHolmes2020}, as shown in Equation \ref{EqGeom}:

\begin{equation} 
\begin{split} 
f_{geom}(N_w=x) = \left(1-p\right)^{x-1}\, p
\label{EqGeom} 
\end{split} 
\end{equation} 

The return period is defined as the expected number of years with no events before a success; for a geometric distribution, this is equal to the inverse of the probability of success, as shown in Equation \ref{EqT}:

\begin{equation} 
\begin{split} 
T=\mathbb{E}[N_w]=
p\sum_{x=1}^\infty x(1-p)^{x-1}= \\ = 
p\frac{1}{p^2} = \frac{1}{p}=\frac{1}{1-\exp{(-\lambda)}}
\label{EqT} 
\end{split} 
\end{equation} 

In the above Equation \ref{EqT} , we used the known series identity $\sum_{k=1}^\infty kr^{k-1}=1/(1-r)^2$ (valid if $-1<r<1$); in our case, we have $r=1-p$. Alternatively, $\mathbb{E}[N_w]$ can be estimated from the cumulative survival function $S_{N_W}(x)=P(N_w>x)=(1-p)^x$ as shown in the following Equation \ref{EqT2} \cite[]{Moore2016}:

\begin{equation} 
\begin{split} 
T=\mathbb{E}[N_w]=\sum_{x=0}^\infty (1-p)^x = \\ =
1+\sum_{x=1}^\infty (1-p)^x= \\ =
1+\sum_{x=1}^\infty \left[\exp{\left(-\lambda \right)}\right]^x = \\ =
1+\sum_{x=1}^\infty \left[\exp{\left(-x \lambda \right)}\right] = \\ =
1+\frac{1-p}{p}=\frac{1}{p}=\frac{1}{1-\exp{(-\lambda)}}
\label{EqT2} 
\end{split} 
\end{equation}

In the above Equation \ref{EqT2}, we considered the known identity of the geometric series $\sum_{n=1}^\infty r^{n}=r/(1-r)$ (valid if $-1<r<1$) with $r=1-p$. A generalization of Equation \ref{EqT2} was proposed by \cite{Cooley2012statistical} for non-stationary processes (Equation \ref{T_NS2}). This expression for $T$ will be later compared to its version under non-stationary conditions in Appendix D, showing how Langbein's formula generalizes for non-stationary processes.

Deriving the relationship between $T$ and $ARI$ is straightforward: by substituting Equation~\ref{EqARI2} into Equation~\ref{EqT}, we obtain Equation~\ref{EqLang}, commonly referred to as Langbein’s formula \cite[]{Langbein1949,Takeuchi1984,Chow49}. Langbein’s formula shows that $T$ approaches 1 asymptotically as $ARI \rightarrow 0$—i.e., $T$ and $ARI$ diverge in the range of frequent events. Conversely, for larger events, $T$ and $ARI$ become increasingly similar. In fact, as $ARI \rightarrow \infty$, their difference asymptotically approaches 0.5 years, as shown in Equation \ref{lim05} \cite[]{WangHolmes2020}:

\begin{equation} 
\begin{split} 
\lim_{ARI\rightarrow\infty}(T-ARI)=
\lim_{ARI\rightarrow\infty}\left\{\left[1-\exp\left(-\frac{1}{ARI}\right)\right]^{-1}-ARI\right\}=
\frac{1}{2}
\label{lim05} 
\end{split} 
\end{equation}

\section{Return period and average recurrence interval under non-stationarity}

\renewcommand{\theequation}{B\arabic{equation}}
\setcounter{equation}{0}

To account for non-stationary conditions, distribution parameters are expressed as functions of time, $t$, either explicitly or through their dependence on climate indices or other covariates that themselves vary over time \cite[]{Coles2001,katz2012statistical,Salas2014}. Consequently, the probability  $p_x=P(Q>q)$ of an annual maximum $Q$ exceeding the fixed reference magnitude $q$ depends on the year $x$ in which the probability is estimated. The return period $T_{NS}$ of the annual event above $q$ under non-stationarity will therefore also depend on the year $t_0$ where it is estimated, i.e., $T_{NS}=T_{NS}(t_0)$. 

Extending the definition of return period under stationary conditions from Appendix A to the non-stationary case, $T_{NS}(t_0)$ is defined as the expected number of uneventful years before the first annual maximum greater than $q$ occurs, modeled using a non-homogeneous geometric distribution ($f_{NHgeom}$; Equation \ref{NHgeom}) starting from year $t_0$ \cite[]{Salas2014}.

\begin{equation}
f_{NHgeom}(N_w=x, t_0) = p_x\,\prod_{k=t_0+1}^{x-1} \left(1 - p_k\right), \quad x = t_0 + 1, t_0 + 2, ..., +\infty
\label{NHgeom}
\end{equation}

In the distribution above, $N_w$ represents the number of years until a success in year $x$, while $p_x$ (or $p_k$) indicates the probability of success—i.e., of an annual maximum greater than the fixed reference $q$—in year $x$ (or $k$). The expression of the expected waiting time—i.e., the return period under non-stationary conditions—is given by Equation \ref{T_NS1}:

\begin{equation}
T_{NS}(t_0)= 
\mathbb{E}\left[N_w|\, t_0 \right] =
\sum_{x=t_0+1}^{+\infty} xp_x\prod_{k=t_0+1}^{x-1} (1-p_k)
\label{T_NS1}
\end{equation}

Here, $x$ and $k$ are used to count the number of years after $t_0$. By performing a series expansion, \cite{Cooley2012statistical} showed that Equation \ref{T_NS1} can also be written as shown in Equation \ref{T_NS2}, which can be regarded as a generalization of Equation \ref{EqT2} under non-stationary conditions:

\begin{equation}
T_{NS}(t_0)=1+\sum_{x=t_0+1}^{+\infty} \, \prod_{k=t_0+1}^{x} (1-p_k)
\label{T_NS2}
\end{equation}

Next, we are going to examine how the average recurrence interval changes under non-stationary conditions. Unlike the stationary case, the rate of exceedances of the reference value $q$ under non-stationary conditions is no longer constant but may vary with time, as a result of changes in the probability of events larger than $q$. This implies that the number of exceedances in a given time interval now follows a non-homogeneous Poisson process—a generalization of the homogeneous Poisson process where the event rate changes over time \cite[]{Coles2001}.
The average recurrence interval $ARI_{NS}$ of events exceeding $q$ under non-stationary conditions is defined as the expected waiting time $\mathbb{E}\left[W\right]$ until the first occurrence, measured from $t_0$. Let $\lambda(t)$ denote the instantaneous rate of annual exceedances of $q$ \cite[also referred to as intensity function;][]{Ross2010,Streit2010,Coles2001} and $N(t_0,\, t_0+w)$ the number of exceedances in the generic interval [$t_0,\ t_0+w$). Then, the expected number of exceedances in that interval is given by Equation \ref{finitInt} \cite[]{Coles2001,Ross2010}:

\begin{equation}
\Lambda(t_0,\, t_0+w)=
\mathbb{E}\left[N\left(t_0, \, t_0+w\right)\right] =
\int_{t_0}^{t_0+w} \lambda(s)ds = 
\int_{0}^w \lambda(t_0+s)ds
\label{finitInt}
\end{equation}

Let $t_1=t_0+w$ denote the time of the first occurrence after $t_0$. The cumulative distribution function of the waiting time $W$, evaluated at $w$, is given by Equation \ref{cdfWaitingTime} \cite[]{Streit2010}:

\begin{equation}
\begin{split}
F_{W,NS}(w,\, t_0) = P(W \leq w) = \\ 
= 1 - P(W > w) = \\ 
= 1 - P(N(t_0,\, t_0 + w) = 0) = \\
= 1 - \exp\left[-\Lambda(t_0,\ t_0 + w)\right] = \\ 
= 1 - \exp\left[-\int_{0}^w \lambda(t_0+s)ds\right]
\label{cdfWaitingTime}
\end{split}
\end{equation}

The expression above uses the fact that the probability of zero exceedances (i.e., $N=0$) in the interval [$t_0, t_0+w$) is equal to $\exp{\left[-\Lambda(t_0,\, t_0+w)\right]}$, if $N$ follows a non-homogeneous Poisson process with intensity $\lambda(t)$ \cite[]{Streit2010}.
Differentiating Equation \ref{cdfWaitingTime} with respect to $w$ yields the probability density function $f_{W,NS}(w,\,t_0)$ of waiting times of the first event after $t_0$ \cite[]{Streit2010,Seno2022}, as shown in Equation \ref{pdfWaitingTime}:

\begin{equation}
f_{W,NS}(w,\, t_0)=\lambda(t_0+w)\,  \exp{\left[-\int_{0}^{w} \lambda(t_0+s)\ ds\right]}
\label{pdfWaitingTime}
\end{equation}

Finally, $ARI_{NS}(t_0)$ is obtained by taking the expected value of $W$, as shown in Equation \ref{ARI_NS1}:

\begin{equation}
\begin{split}
ARI_{NS}(t_0)=\mathbb{E}\left[W| \, t_0\right] = \int_{0}^{\infty} w\, \lambda(t_0+w)\  \exp{\left[ -\int_{0}^{w}\lambda(t_0+s)\ ds \right]}\ dw
\label{ARI_NS1}
\end{split}
\end{equation}

As we already illustrated for the stationary case (Appendix A), a simpler expression for the non-stationary $ARI_{NS}$ can be derived integrating Equation \ref{ARI_NS1} by parts, obtaining the following Equation \ref{ARI_NS2}:

\begin{equation}
\begin{split}
ARI_{NS}(t_0)=\int_0^\infty w f_{W,NS}(w,\, t_0)\, dw = \\ 
=  \left[ -w(1 - F_{W,NS}(w,\, t_0)) \right]_0^\infty 
+ \int_0^\infty (1 - F_{W,NS}(w,\,t_0))\, dw = \\ 
= \int_0^\infty \exp \left[ -\int_{0}^{w}\lambda(t_0+s)\ ds \right]dw = \\
=\int_0^\infty \exp{\left[-\Lambda(t_0,\, t_0+w) \right] dw}
\end{split}
\label{ARI_NS2}
\end{equation}

Equation \ref{ARI_NS2} represents a generalization of Equation \ref{EqARI2} to the non-stationary case; in both Equations, the average recurrence interval is given as the integral of the survival function, $S_W(w)=1-F_W(w)$ (or $S_{W,NS}(w, \, t_0)=1-F_{W,NS}(w, \, t_0)$ for the non-stationary case).

We conclude this section by addressing the different types of timelines considered by the two formulations for $T_{NS}(t_0)$ and $ARI_{NS}(t_0)$—discrete in one case, continuous in the other. Since $T_{NS}(t_0)$ represents the expected number of years without events prior to an annual maximum exceeding a given reference magnitude, Equations \ref{T_NS1} and \ref{T_NS2} are formulated over a discretized  timeline—i.e., $x=t_0+1,$ $t_0+2$, ..., —where temporal variability in the distributional characteristics of the process is considered at the annual scale. This approach aligns with numerous applied examples in the literature \cite[]{Salas2014,katz2012statistical,Coles2001}. In contrast, the estimation of $ARI$ as the expected waiting time to the next exceedance assumes a continuous time framework. To ensure consistency between these formulations, we adopt values of the probability $p_x$ averaged over the one-year interval [$x-1$, $x$), to obtain the exceedance probability for year $x$. By considering the instantaneous exceedance rate $\lambda(t)$, reflecting temporal changes in the distributional characteristics of the process, and substituting $p_k=1-\exp{\left(-\int_{k-1}^k \lambda(t_0+s) ds\right)}$ into Equations \ref{T_NS1} and \ref{T_NS2}, we obtain:

\begin{equation}
\begin{split}
T_{NS}(t_0)=\sum_{x=t_0+1}^{+\infty} xp_x\prod_{k=t_0+1}^{x-1} (1-p_k) = \\ = 
\sum_{x=t_0+1}^{+\infty} x \left[1-\exp{\left(-\int_{x-1}^x  \lambda(s) ds\right)}\right]
\prod_{k=t_0+1}^{x-1}\exp{\left[-\int_{k-1}^{k}\lambda(s)ds \right]} = \\ 
= \sum_{x=t_0+1}^{+\infty} x\left[1-\exp{\left(-\int_{x-1}^x  \lambda(s) ds\right)}\right]
\exp{\left[-\int_{t_0}^{x-1}\lambda(s)ds \right]} = \\ =
\sum_{x=t_0+1}^{+\infty} x\left(1-\exp{\left[-\Lambda(x-1,\,x) \right]}\right)\exp{\left[-\Lambda(t_0,\, x-1) \right]}
\label{T_NS1_stepInt}
\end{split}
\end{equation}

\begin{equation}
\begin{split}
T_{NS}(t_0)=1+\sum_{x=t_0+1}^{+\infty} \prod_{k=t_0+1}^{x} (1-p_k) = \\ =
1+\sum_{x=t_0+1}^{+\infty}\prod_{k=t_0+1}^{x}\exp{\left[-\int_{k-1}^{k}\lambda(s)ds \right]} = \\
= 1+\sum_{x=t_0+1}^{+\infty}\exp{\left[-\int_{t_0}^{x}\lambda(s)ds \right]} = \\ = 
1+\sum_{x=t_0+1}^{+\infty}\exp{\left[-\Lambda(t_0,\, x)\right]}
\label{T_NS2_stepInt}
\end{split}
\end{equation}

In deriving the final forms of Equations \ref{T_NS1_stepInt} and \ref{T_NS2_stepInt}, we applied the exponential addition rule and the additivity of definite integrals. These expressions represent the return period in terms of the cumulative intensity function $\Lambda(t)$, and will be instrumental in demonstrating the validity of Langbein’s formula under non-stationary conditions (Appendix D).

\section{Details about the example in Figure \ref{FigNS}}

\renewcommand{\theequation}{C\arabic{equation}}
\setcounter{equation}{0}

Drawing inspiration from selected examples in \cite{katz2012statistical}, we defined two arbitrary non-stationary processes each described by a Generalized Extreme Value (GEV) distribution—parametrized as in \cite{katz2012statistical}—with time-varying location, scale and shape parameters (i.e., $\mu_g(t)$, $\sigma_g(t)$, and $\xi_g(t)$) according to Equations \ref{EqGEVparams_t} with coefficients $\mu_{g,0}$, $\mu_{g,1}$, $\sigma_{g,0}$, $\sigma_{g,1}$, and $\xi_g$:

\begin{equation}
\mu_g(t)=\mu_{g,0}+\mu_{g,1}\ t, \quad \ln \sigma_g(t)=\sigma_{g,0}+\sigma_{g,1}\ t, \quad \xi_g(t)=\xi_g
\label{EqGEVparams_t}
\end{equation}

By adopting the above Equations \ref{EqGEVparams_t}, we introduced a linear trend in the location parameter $\mu_g(t)$—characterized by slope $\mu_{g,1}$ and an initial value $\mu_{g,0}$ at time $t_0 = 0$—as well as an exponential growth in the scale parameter $\sigma_g(t)$ (assuming $\sigma_{g,1} > 0$). Regarding the shape parameter, it is assumed constant over time; this parameter is already challenging to estimate in stationary extreme value models, due to its strong sensitivity to sampling variability, therefore, modeling $\xi$ as a function of time would be unrealistic \cite[]{Coles2001}. In these expressions, time $t$ is continuous and expressed in years. Trends in $\mu_g(t)$ and $\sigma_g(t)$ are assumed to occur over a 15-year period starting from $t_0=0$, after which the distribution's properties are assumed to remain constant and equal to their values at year 15. Table \ref{TabNSexamplesGEV} summarizes the values of the coefficients adopted for the two non-stationary GEV models and also provides the value of the threshold $\mu_p$ for the corresponding non-stationary GP model.

\begin{table}[h!]
\caption{Model coefficients for the two non-stationary GEV distributions shown in Figure \ref{FigNS}}
\centering
\begin{tabular}{c c c c c c c}
\hline
Example & $\mu_{g,0}$ & $\mu_{g,1}$ & $\sigma_{g,0}$ & $\sigma_{g,1}$ & $\xi_g$ & $\mu_p$ \\
\hline
\#1  & 109.43 & 0.86 & 2 & 0.04 & 0.3 & 100.0  \\
\#2  & 152.03 & 0.76 & 3.5 & 0.05 & 0.03 & 100.0\\
\hline
\end{tabular}
\label{TabNSexamplesGEV}
\end{table}

Under the assumption that the number of threshold exceedances follows a non-homogeneous Poisson process (NHPP) with instantaneous rate of annual exceedances $\lambda(t)$ (see also Appendix B), the time-varying parameters of the non-stationary Generalized Pareto (GP) distribution of exceedance magnitudes—corresponding to the non-stationary GEV distribution of annual maxima—can be obtained at any time $t$ from the duality between the two distributions \cite[]{katz2012statistical,WangHolmes2020}. Specifically, we considered the following Equations \ref{EqGPparams_t} \cite[]{katz2012statistical} to derive the non-stationary GP model.

\begin{equation}
\xi_p=\xi_g, \quad \lambda(t)=\left(1+\xi\frac{\mu_p-\mu_g(t)}{\sigma_g(t)}\right)^{-\frac{1}{\xi}}, \quad \sigma_p(t)=\sigma_g(t)+\xi\left[\mu_p-\mu_g(t)\right]
\label{EqGPparams_t}
\end{equation}

Here, $\lambda(t)$ represents the time-varying rate of threshold exceedances. The initial rate $\lambda(t_0)$ was set to 5 in both examples shown in Figure \ref{FigNS}. For the parameter values adopted in the non-stationary models (see Table \ref{TabNSexamplesGEV}), $\lambda(t)$ gradually increases to 10 in the example in Figure \ref{FigNS}a, while it decreases to 2.5 in the example in Figure \ref{FigNS}b. The time-varying rate of exceedances $\lambda_Q(t)$ for magnitudes $Q$ above the threshold of the PD series (shown in panels a2 and b2 in Figure \ref{FigNS}) was obtained using Equation \ref{EqLambda_q_t} \cite[]{katz2012statistical}:

\begin{equation}
\lambda_Q(t)=\left(1+\xi\frac{Q-\mu_g(t)}{\sigma_g(t)}\right)^{-\frac{1}{\xi}}
\label{EqLambda_q_t}
\end{equation}

In the two examples shown in Figure \ref{FigNS}, we considered events with $ARI_0$—that is, the $ARI$ at time $t_0$ if the process were stationary—of 0.2, 0.5, 1, 2, 5, 10, 20, 50, and 100 years. Note that the 0.2-year event—corresponding to an expected exceedance rate of five times per year—also served as the threshold for the non-stationary GP model in both examples.

\section{Analytical demonstration of the validity of Langbein's formula under non-stationarity}

\renewcommand{\theequation}{D\arabic{equation}}
\setcounter{equation}{0}

We observed in Section 2 and Figure \ref{FigNS} that $T$-$ARI$ pairs obtained under non-stationary conditions (denoted as $T_{NS}(t_0)$-$ARI_{NS}(t_0)$), all estimated at the same time $t_0$ for a range of events with a given (stationary) $ARI_0$ at year $t_0$, lie on Langbein's theoretical curve (Equation \ref{EqLang}), even though they do not occupy the same position along the curve as they would under stationary conditions. Because the two considered examples of non-stationary processes were arbitrary, this finding can be regarded as an empirical evidence that Langbein's formula holds true under non-stationarity, with the caveat that $T_{NS}$ and $ARI_{NS}$ must be estimated using appropriate formulations (Equations \ref{T_NS} and \ref{ARI_NS}). In this appendix, we demonstrate more rigorously the formula's validity under non-stationarity.

Langbein's equation was originally derived under the assumption that the number of exceedances follows a Poisson process with constant expected number of exceedances \cite[i.e., a homogeneous Poisson process;][]{Takeuchi1984}. However, in non-stationary processes, the probability of exceedance and hence the rate of exceedance of a predefined threshold both change with time. If we indicate as $\lambda(t)$ the annual rate of exceedances at time $t$, the expected number of exceedances in the interval [$0$, $t$) is given by the following Equation \ref{EqLambda_t} \cite[]{Streit2010}:

\begin{equation}
\Lambda(t)= \int_{0}^{t} \lambda(s)ds
\label{EqLambda_t}
\end{equation}

Note that, in contrast with Equation \ref{finitInt}, here we do not specify the lower bound in the argument of $\Lambda(t)$ when it is 0. For simplicity, in this appendix we will consider 0 as the initial time $t_0$.

Because of the time-varying $\Lambda(t)$, the number $N(t)$ of exceedances up to time $t$ follows a non-homogeneous Poisson process (NHPP) with intensity $\lambda(t)$, as shown by Equation \ref{EqNHPP}:

\begin{equation}
P(N(t)=k)= \frac{[\Lambda(t)]^k}{k!}\ \exp{\left(-\Lambda(t)\right)}
\label{EqNHPP}
\end{equation}

The latter is a generalization of the homogeneous Poisson process. We will now show that a NHPP can be regarded as a homogeneous Poisson process (HPP) in a transformed time scale \cite[]{kingman1992poisson}. Let us define the time transformation given by Equation \ref{Eqt2tau}, where $\lambda_0$ is the initial excess rate at time $t_0$ where we estimate $T_{NS}$ and $ARI_{NS}$; we are going to show that $\lambda_0$ is the constant rate of the HPP in $\tau$-space for a unit interval.

\begin{equation}
\begin{split}
\tau = \frac{1}{\lambda_0}\Lambda(t)= \frac{1}{\lambda_0}\int_0^t \lambda(s)\ ds
\label{Eqt2tau}
\end{split}
\end{equation}

The product of transformed time by this constant is equal to the expected number of events up to the original time $t$ (i.e., $\Lambda(t)=\lambda_0\, \tau$). The transformed time scale $\tau$ therefore expands for larger values of $\lambda(t)$ in the original time scale $t$ and contracts for smaller values of $\lambda(t)$. Since $\lambda(t)>0$ at any $t$, $\Lambda(t)$ increases monotonically with $t$, ensuring that its inverse,  $t=\Lambda^{-1}(\lambda_0\,\tau)$, exists \cite[]{kingman1992poisson}. By defining the new counting process in the scaled time $\tau$, $M(\lambda_0\, \tau)=N(\Lambda^{-1}(\lambda_0\,\tau))$, Equation \ref{EqNHPP} can be reformulated into Equation \ref{EqNHPP2HPP}:

\begin{equation}
\begin{split}
P(M(\lambda_0\, \tau) = k) = P(N(\Lambda^{-1}(\lambda_0\, \tau))=k) = \\ 
=\frac{[\Lambda(\Lambda^{-1} (\lambda_0\, \tau))]^k}{k!}\,\exp(-\Lambda(\Lambda^{-1} (\lambda_0\, \tau))= \\ =\frac{(\lambda_0\, \tau)^k}{k!}\,\exp(-\lambda_0\, \tau)
\label{EqNHPP2HPP}
\end{split}
\end{equation}

The expression above is a HPP with mean $\lambda_0\, \tau$. Following the same steps shown in Appendix A, we can define $ARI_{\tau}$ in the transformed time $\tau$ as the expected value of the waiting time associated with that HPP (Equation \ref{EqARI_tau}), and the corresponding $T_{\tau}$ as the expected number of failures before a success in a Bernoulli trial with probability of success $p=1-\exp(\lambda_0)$ (Equation \ref{EqT_tau}).

\begin{equation}
\begin{split}
ARI_{\tau}=\frac{1}{\lambda_0}
\label{EqARI_tau}
\end{split}
\end{equation}

\begin{equation}
\begin{split}
T_{\tau}=\frac{1}{1-\exp\left(-\lambda_0\right)}
\label{EqT_tau}
\end{split}
\end{equation}

Substituting Equation \ref{EqARI_tau}  into Equation \ref{EqT_tau}, we obtain Langbein's formula in the transformed time, as shown in Equation \ref{EqLang_tau}:

\begin{equation}
\begin{split}
\displaystyle T_{\tau}=\frac{1}{1-\exp\left(\displaystyle -\frac{\displaystyle 1}{\displaystyle ARI_{\tau}}\right)}
\label{EqLang_tau}
\end{split}
\end{equation}

We just showed that $T_{\tau}$-$ARI_{\tau}$ pairs in the transformed time scale $\tau$ (Equation \ref{Eqt2tau}) satisfy Langbein's formula. Switching to the original time scale $t$, $ARI_{NS}$ (Equation \ref{ARI_NS2}) can be expressed in terms of $\tau$ by substituting the reparameterization equations \ref{EqTransforms}—associated with the time transformation defined in Equation \ref{Eqt2tau}—into Equation \ref{ARI_NS2}, obtaining Equation \ref{EqARI_NS2tau}. For simplicity, we considered $t_0=0$ in our derivations.

\begin{equation}
\Lambda(t)=\lambda_0\,\tau, \quad t=\Lambda^{-1}(\lambda_0\,\tau), \quad \frac{d\tau}{dt}=\frac{1}{\lambda_0}\lambda(t), \quad \frac{dt}{d\tau}=\frac{\lambda_0}{\lambda\left[\Lambda^{-1}(\lambda_0\,\tau)\right]}
\label{EqTransforms}
\end{equation}

\begin{equation}
\begin{split}
ARI_{NS}(t_0=0) = \int_0^\infty \exp{\left[-\Lambda(t)\right]}dt = \\ =\int_0^\infty \exp{\left(-\lambda_0 \tau\right)\, \frac{\lambda_0}{\lambda(\Lambda^{-1}(\lambda_0 \tau))}\ d\tau}
\label{EqARI_NS2tau}
\end{split}
\end{equation}

This new expression for $ARI_{NS}$ closely recalls the one for $ARI$ in stationary conditions provided in Equation \ref{EqARI2}—which calculates $ARI$ as the integral of the survival function $S_W(w)=$ $1-F_W(w)$=$\exp{\left(-\lambda w \right)}$, with the difference that now the survival function is weighted by the time-dilation factor $\lambda_0\left[\lambda(\Lambda^{-1}(\lambda_0 \tau))\right]^{-1}$, causing $ARI_{NS}$ to either contract or expand depending on whether $\lambda(t)$ increases or decreases with respect to $\lambda_0$ in the interval of integration, respectively. This theoretical behavior is in agreement with our observations in the examples provided in Figure \ref{FigNS}. It is easy to show that Equation \ref{EqARI_NS2tau} represents a generalization of Equation \ref{EqARI2}; assuming $\lambda(t)=\lambda$ (i.e., the threshold-exceedance process is stationary) into Equation \ref{EqARI_NS2tau} leads to obtaining the same result as in Equation \ref{EqARI2}, as shown below:

\begin{equation}
\begin{split}
ARI=\int_0^\infty \exp{\left(-\lambda_0 \tau\right)\,\frac{\lambda_0}{\lambda}\ d\tau}=\frac{\lambda_0}{\lambda}\left[-\frac{1}{\lambda_0}\exp{(-\lambda_0 \tau)}\right]_0^\infty=\frac{\lambda_0}{\lambda}\cdot\frac{1}{\lambda_0}=\frac{1}{\lambda}
\label{EqARI_NStau_constLambda}
\end{split}
\end{equation}

Concerning $T_{NS}$, by substituting the time-transformation equations \ref{EqTransforms} into the expression for $T_{NS}$ provided by Equation \ref{T_NS2_stepInt}, we obtain:

\begin{equation}
\begin{split}
T_{NS}(t_0=0)= 1+\sum_{x=1}^{x_{\max}}\exp{\left[-\Lambda(x)\right]} = 1+\sum_{x=1}^{x_{\max}}\exp{\left[-\tau_x \lambda_0 \right]}, \quad \tau_x=\frac{1}{\lambda_0}\int_0^x \lambda(s) ds
\label{EqT_NS2tau}
\end{split}
\end{equation}

In the Equation above, we introduced $\tau_x$, which corresponds to the generic year $x$ in the transformed time scale, according to Equation \ref{Eqt2tau}. The above Equation \ref{EqT_NS2tau} is very similar to the expression of $T$ in stationary conditions shown in Equation \ref{EqT2}, the only difference being the presence of the time-dilating coefficient $\tau_x$ in the argument of the exponential function, in place of the integer and uniformly spaced variable $x$ (counting the years). It is straightforward to show that, if we assume stationary conditions (i.e., if we assume $\lambda(t)=\lambda$), \ref{EqT_NS2tau} simplifies into \ref{EqT2}, as $\tau_x$ in that case would be equal to $x\, \lambda/\lambda_0$. When $\tau_x>x$ (i.e., when the expected number of exceedances in the interval [0, $x$) is greater than if we had stationary conditions), the contribution of the corresponding addends in the summation is smaller, as compared with a stationary process with rate $\lambda_0$ constant over time (because the argument in the exponential function has negative sign), resulting in a contraction of $T_{NS}$ with respect to $T$. This occurs for values of the instantaneous rate $\lambda(t)$ larger than $\lambda_0$, which causes time in the transformed scale to accelerate (i.e., $\tau_x>x$). In a similar way, when the rate of exceedances $\lambda(t)$ decreases, time in the transformed time scale $\tau$ slows down (i.e., we observe $\tau_x<x$), causing a dilation of $T_{NS}$ as compared with its stationary counterpart. This theoretical behavior is in agreement with our observations in the examples provided in Figure \ref{FigNS}.

So far, we have showed that $T_{\tau}$ and $ARI_{\tau}$ in the transformed scale satisfy Langbein's formula (as a result of the fact that exceedances follow a HPP in the transformed scale), and that this transformation provides an analytical framework to measure the effects of non-stationarity on the concurrent expansion or contraction of  $T_{NS}$ and $ARI_{NS}$ with respect to the stationary case in the original time scale. However, we have not yet shown analytically that $T_{NS}$ and $ARI_{NS}$ in the original time scale lie on Langbein's theoretical curve. To address this, we will now consider the expressions of $T_{NS}$ and $ARI_{NS}$ in $t$ given by Equations \ref{T_NS2_stepInt} and \ref{ARI_NS2}; for simplicity,  we will substitute the expression of the survival function, $S_{W,NS}(t)=\exp \left[ -\Lambda(t) \right]$, obtaining:

\begin{equation}
\begin{split}
ARI_{NS}(t_0=0) = \int_0^\infty \exp{\left[ -\Lambda(t)\right]}dt = \\ 
= \int_0^\infty S_W(t)dt
\label{EqARI_Sw}
\end{split}
\end{equation}

\begin{equation}
\begin{split}
T_{NS}(t_0=0) = 1+\sum_{x=1}^\infty \exp \left[ -\Lambda(x)\right] = \\
= 1+\sum_{x=1}^\infty S_W(x)
\label{EqT_Sw}
\end{split}
\end{equation}

Since $S_W(t)$ is positive non-increasing, we can provide the following non-negative lower and upper bounds on the sum:

\begin{equation}
\begin{split}
\int_1^\infty S_W(t)dt = \sum_{k=1}^\infty \int_k^{k+1} S_W(t)dt \le \sum_{k=1}^\infty S_W(k) \le S_W(1) + \int_1^\infty S_W(t)dt
\label{lbub}
\end{split}
\end{equation}

The first side of the inequality in Equation \ref{lbub} is valid for any positive non-increasing $S_W(t)$ because in that case $S_W(k) \ge \int_k^{k+1}S_W(t)dt$—i.e., the integral over [$k$, $k+1$] is smaller than or equal to the area $S_W(k) \cdot 1$ of the rectangle of height $S_W(k) \ge S_W(k+1)$ and width 1. The upper bound is also always valid for any positive non-increasing $S_W(t)$, because $S_W(k) \le \int_{k-1}^k S_W(t)dt$—i.e., this time we are comparing the integral and the area of the rectangle with the lower height—and therefore $\sum_{k=1}^\infty S_W(t) = S_W(1)+\sum_{k=2}^\infty S_W(k) \le S_W(1) + \int_1^\infty S_W(t) dt$. By considering that $S_W(1) \le \int_0^1 S_W(t)$ (i.e., the integral over [$0$, $1$] is larger than or equal to the area $S_W(1) \cdot 1$ of the rectangle of height $S_W(1) \le S_W(0)$), it can be immediately noticed that $ARI_{NS}$ is in turn an upper bound for the upper bound in the expression above:

\begin{equation}
\begin{split}
S_W(1) + \int_1^\infty S_W(t) \, dt \le \int_0^1 S_W(t) \, dt + \int_1^\infty S_W(t) \, dt = {ARI_{NS}}
\label{ARIub}
\end{split}
\end{equation}

The lower bound in Equation \ref{lbub} is non-negative since it is obtained as the integral of a non-negative function. In conclusion, the term $\sum_{x=1}^\infty S_W(x)$ in the expression of $T_{NS}$ is bounded by the following inequalities:

\begin{equation}
\begin{split}
0 \le \sum_{k=1}^\infty S_W(k) \le ARI_{NS}
\label{lbub2}
\end{split}
\end{equation}

As $ARI \rightarrow 0$, by the Squeeze Theorem, $\sum_{k=1}^\infty S_W(k) \rightarrow 0$ as well. Therefore, from the expression of $T_{NS}$ given in Equation~\ref{EqT_Sw}, we have $T_{NS} \rightarrow 1$ when $ARI_{NS} \rightarrow 0$. This finding indicates that $T_{NS}$–$ARI_{NS}$ pairs lie on the theoretical curve defined by Equation~\ref{EqNSLang} in the regime of small $ARI$s, with a displacement along the curve from the stationary $T$–$ARI$ counterpart that depends on the rate of increase (or decrease) of $\lambda(t)$, as illustrated earlier by deriving Equations~\ref{EqARI_NS2tau} and~\ref{EqT_NS2tau}.

We will now demonstrate that, for large $ARI_{NS}$, $T_{NS} \approx ARI_{NS} +0.5$, indicating that $T_{NS}$-$ARI_{NS}$ pairs lie on the theoretical curve defined by Equation \ref{EqNSLang} even in the range of the larger events (see Appendix A, Equation \ref{lim05}). To achieve this, we use the Euler-Maclaurin summation, provided below, (Equation \ref{EulerMaclaurin}) to approximate the sum of the survival function, $\sum_{x=1}^\infty S_W(x)$.

\begin{equation}
\begin{split}
\sum_{k=a}^{b} f(k) = \int_a^b f(x)\,dx + \frac{f(a) + f(b)}{2} + \sum_{n=1}^{m} \frac{B_{2n}}{(2n)!} \left( f^{(2n-1)}(b) - f^{(2n-1)}(a) \right) + R_m, \quad b>a
\label{EulerMaclaurin}
\end{split}
\end{equation}

In the Equation above, $f$ is a generic function, $B_{2n}$ denote even-order Bernoulli numbers (e.g., Bernoulli numbers $B_n$ in the range from $n=0$ to $n=7$ are: -1/2, 1/6, 0, -1/30, 0, 1/42, 0, respectively), $R_m$ indicates the $m$-order truncation term, and $f^{2n-1}$ represent odd-order derivatives of $f$.
The Euler-Maclaurin summation formula (Equation \ref{EulerMaclaurin}) is stated for finite intervals [$a,b$] but our summation extends to infinity; nevertheless, since $S_W(x)$ has a horizontal asymptote at 0, as $x$ goes to infinity, both $S_W(x)$ and its derivatives become negligible for large values of $x$. This means that, if we set $b$ sufficiently large, we can still use Equation \ref{EulerMaclaurin} to approximate the sum $\sum_{x=1}^\infty S_W(x)$ by neglecting the terms evaluated at $b$. As we are interested in the range of Langbein's curve associated with the larger, rarer events, the instantaneous exceedance rate $\lambda(t)$ is small and the exceedance rate $\Lambda(x)$ over a finite interval [$0$, $x$] grows slowly with time, hence, the derivatives of the survival function $S_W(x)=\exp\left[- \Lambda(x)\right]$ are also negligible; for example, the first-order derivative would be $S_W'(x)=-\lambda(x)\exp \left[ -\Lambda(x)\right]$, which is significantly smaller than $S_W(x)$, since $\lambda(t)$ is close to 0 in the range of rarer events, regardless of its variability. By truncating after the zero-order terms, we obtain the following approximation of $\sum_{x=1}^\infty S_W(x)$ (Equation \ref{approxSumS}):

\begin{equation}
\begin{split}
\sum_{x=1}^\infty S_W(x) \approx \int_1^\infty S_W(t) \, dt + \frac{S_W(1)}{2} 
\label{approxSumS}
\end{split}
\end{equation}

By substituting the expression for $ARI_{NS}$ from Equation \ref{EqARI_Sw} into Equation \ref{approxSumS}, we obtain:

\begin{equation}
\begin{split}
\sum_{x=1}^\infty S_W(x) \approx ARI_{NS}-\int_0^1 S_W(t) \, dt + \frac{S_W(1)}{2} \approx \\ \approx
ARI_{NS}-\int_0^1 \left[1-\Lambda(t)\right] \, dt + \frac{1-\Lambda(1)}{2} \approx \\\approx
ARI_{NS}-1 + \frac{1}{2} 
\label{approxSumS_withARI}
\end{split}
\end{equation}

In the second line of Equation \ref{approxSumS_withARI}, we considered the first-order Taylor approximation for the exponential function, $S_W(t)=\exp \left[ -\Lambda(t)\right]$ $\approx 1-\Lambda(t)$. This approximation is valid under the assumption that $\Lambda(t)\ll 1$, which holds in our case due to the rarity of threshold exceedances and the short duration—only one year—of the interval [0, 1] where $\Lambda(t)$ is calculated. Then, to obtain the last line, we substituted $\Lambda(t)\approx 0$ and $\Lambda(1)\approx 0$, both valid in the range of rare events and for the short, 1-year interval where $\Lambda(t)$ is obtained.
We are now ready to demonstrate that $T_{NS}\approx ARI_{NS}+1/2$ for large $ARI$, as it is for the stationary case (Equation \ref{lim05}). By substituting Equation \ref{approxSumS_withARI} into the expression for $T_{NS}$ given in Equation \ref{EqT_Sw}, we obtain:

\begin{equation}
\begin{split}
T_{NS}  =  1+\sum_{x=1}^\infty S_W(x) \approx \\
\approx
1 + ARI_{NS} -1 + \frac{1}{2}= \\ 
= ARI_{NS} + \frac{1}{2}
\label{demonstrationLangNS}
\end{split}
\end{equation}

Equation \ref{demonstrationLangNS} indicates that $T_{NS}$-$ARI_{NS}$ pairs lie on the theoretical curve defined by Equation \ref{EqNSLang}, also for large $ARI$s, with a displacement along the curve from the stationary $T$-$ARI$ counterpart that depends on the rate of change of $\lambda(t)$ relative to $\lambda(t_0=0)=\lambda_0$ at the time of estimating $T_{NS}$ and $ARI_{NS}$, as governed by Equations \ref{EqARI_NS2tau} and \ref{EqT_NS2tau}. This finding confirms that Equation \ref{EqLang} can be generalized into Equation \ref{EqNSLang} under non-stationarity even when considering large events.

\section{Return period for processes exhibiting over-dispersion}

\renewcommand{\theequation}{E\arabic{equation}}
\setcounter{equation}{0}

Let the number of exceedances in a year be a random variable $N$ describing the number of failures before collecting a total of $r$ successes (with $r>0$) in a Bernoulli experiment with probability of success $p$ ($0<p<1$), considering $k+r$ trials. Then, $N$ follows a Negative Binomial distribution and its probability is given by Equation \ref{NegBin} \cite[]{Bezak2014}:

\begin{equation}
P(N = k) = \binom{k + r - 1}{k} \, (1 - p)^k \, p^r , \quad k = 0, 1, 2, \dots
\label{NegBin}
\end{equation}

In the above Equation \ref{NegBin}, $k \ge 0$ denotes the realization of the number of failures before reaching the last, $r^{th}$ success. It follows that $k+r-1$ is the number of Bernoulli trials before reaching the $r^{th}$ success, (or equivalently, $k+r$ is the total number of trials including the last, successful one). Both $p$ and $r$ depend on the mean and variance of the number of exceedances, if the distribution is used as a count model. Lastly, $\binom{k+r-1}{k}$ is the binomial coefficient, representing the number of all possible ways to arrange $k$ failures and $r-1$ successes.

The mean of the number $N$ of exceedances associated with a fixed threshold $u$, $\mathbb{E}\left[N\right]=\lambda_u$, and its variance $\operatorname{Var}\left[N\right]=V_u$—or equivalently, the mean and variance of the number of failures in the sequence of $k+r$ Bernoulli trials—are given by Equations \ref{NegBinE} and \ref{NegBinV} \cite[]{Bezak2014}:

\begin{equation}
\mathbb{E}\left[N\right]=\lambda_u=\frac{r(1-p)}{p}
\label{NegBinE}
\end{equation}

\begin{equation}
\operatorname{Var}\left[N\right]=V_u=\frac{r(1-p)}{p^2}
\label{NegBinV}
\end{equation}

Note that $\operatorname{Var}\left[N\right]>\mathbb{E}\left[N\right]$ for any $r>0$, $0<p<1$, which is a characteristic property of the Negative Binomial distribution. By rearranging the two Equations \ref{NegBinE}  and \ref{NegBinV}, we obtain the expressions for $p$ and $r$ given in Equations \ref{NegBinP} and \ref{NegBinR}:

\begin{equation}
p=\frac{\lambda_u}{V_u}
\label{NegBinP}
\end{equation}

\begin{equation}
r=\frac{\lambda_u^2}{V_u-\lambda_u}
\label{NegBinR}
\end{equation}

Equations \ref{NegBinP} and \ref{NegBinR} can be used to calibrate the count model to the series of exceedance counts for a given threshold $u$, by using the sample mean and sample variance of excess counts. We now focus on characterizing how the variance of the number of exceedances changes with the threshold—or, equivalently, with the exceedance rate. Let $\lambda$ and $V$ denote the expected value and variance, respectively, of the number of exceedances for an arbitrary threshold. By rearranging Equation \ref{NegBinR} and considering that $r$ does not change with the threshold (find a demonstration of this in Appendix G), we obtain the expression of $V$ given in Equation \ref{NegBinV_lambda}:

\begin{equation}
V = \lambda + \frac{\lambda^2}{r} = \lambda \left(1+\frac{\lambda}{r}\right)
\label{NegBinV_lambda}
\end{equation}

The term $1/r>0$ in Equation \ref{NegBinV_lambda} acts as dispersion parameter, controlling how much larger the variance $V$ is relative to the mean number of exceedances $\lambda$, through the additive term $\lambda^2/r$. Notably, $V$ tends toward $\lambda$ (i.e., equi-dispersion) as $r\rightarrow+\infty$ \cite[]{Zhou2015}.

We can now derive the return period $T$ under over-dispersion. From the expression of the Negative Binomial distribution (Equation \ref{NegBin}) and considering Equation \ref{NegBinP}, the probability of no events in a given year $t$—i.e., in the interval [$t-1,~ t$)—, and the probability of at least one event in that year are given by Equations \ref{pX_t_NB0} and \ref{pX_t_NB1}, respectively:

\begin{equation}
P \left(\left[ N_{t}-N_{t-1} \right]=0 \right) =
p^r= 
\left( \frac{\lambda}{V} \right)^r =
\left(1+\lambda/r\right)^{-r}
\label{pX_t_NB0}
\end{equation}

\begin{equation}
P \left(\left[ N_{t}-N_{t-1} \right]>0 \right) =
1-P(\left[ N_{t}-N_{t-1} \right]=0)=
1- \left( \frac{\lambda}{V} \right)^r =
1-\left(1+\lambda/r\right)^{-r}
\label{pX_t_NB1}
\end{equation}

Note that the right-hand side of Equations \ref{pX_t_NB0} and \ref{pX_t_NB1} was obtained by replacing Equation \ref{NegBinV_lambda}. By definition, the return period is equal to the inverse of the probability of observing an annual maximum (Equation \ref{EqT}); by substituting Equation \ref{pX_t_NB1} into Equation \ref{EqT}, we obtain the return period for processes exhibiting over-dispersion, given in Equation \ref{EqT_NB}:

\begin{equation}
\begin{split}
T=\frac{1}{1- \left(\lambda /V \right)^r} = \\ =
\frac{1}{1- \left(1+\lambda/r\right)^{-r}}
\label{EqT_NB}
\end{split}
\end{equation}

Equation \ref{EqT_NB} can be regarded as a generalization of Equation \ref{EqT}. Taking the limit as $V \rightarrow \lambda$ (which occurs when $r \rightarrow +\infty$, see Equation \ref{NegBinV_lambda}) yields the return period for a homogeneous Poisson process (i.e., under equi-dispersion; Equation \ref{EqT}), as expected when $V \approx \lambda$. Specifically, by introducing $x = \lambda / r$ (so that $x \rightarrow 0$ as $r \rightarrow \infty$), we can write:

\begin{equation}
\begin{split}
\lim_{r \rightarrow +\infty}{\left(1+\lambda/r\right)^{-r}}=
\lim_{x \rightarrow 0}{\left( 1+x \right)^{-\lambda/x }} = \exp \left(-\lambda \right)
\label{EqLim}
\end{split}
\end{equation}

Substituting Equation \ref{EqLim} into Equation \ref{EqT_NB} yields exactly the same expression for $T$ as in Equation \ref{EqT}, confirming that the latter is a limiting case of Equation \ref{EqT_NB} for $V \rightarrow \lambda$. In the equation above, we used the standard limit $\lim_{x \rightarrow 0} (1 + x)^{1/x} = e$.

Equation \ref{EqT_NB}, derived using a Negative Binomial count model—accounting for over-dispersion in the number of exceedances—converges to the equi-dispersed case as $V \rightarrow \lambda$. In Appendix F, we demonstrate that this general form also holds for under-dispersed processes; see also Equation \ref{EqT_Gen_psi} for a unified formulation that accommodates both over- and under-dispersion in excess counts.

\section{Return period for processes exhibiting under-dispersion}

\renewcommand{\theequation}{F\arabic{equation}}
\setcounter{equation}{0}

Let the number of exceedances in a year be a random variable $N$ describing the number of successes in $\gamma$ trials (with $\gamma>0$) of a Bernoulli experiment with probability of success $p$ ($0<p<1$). Therefore, $N$ follows a Binomial distribution and its probability is given by Equation \ref{Bin} \cite[]{Bezak2014}:

\begin{equation}
P(N = k) = \binom{\gamma}{k} \, (1 - p)^{\gamma-k} \, p^k , \quad k = 0, 1, 2, \dots
\label{Bin}
\end{equation}


In the above Equation \ref{Bin}, $k \ge 0$ is the realization of the number of successes in $\gamma$ Bernoulli trials, while $\binom{\gamma}{k}$ is the binomial coefficient representing the number of all possible ways to arrange $k$ successes in $\gamma$ trials. If the Binomial distribution is used as a count model, then both $\gamma$ and $p$ depend on the mean and variance of exceedance counts. 

For a fixed threshold $u$, the mean number of exceedances $\mathbb{E}\left[N\right]=\lambda_u$ and its variance $Var\left[N\right]=V_u$ (or, equivalently, the mean and variance of the number of successes in the sequence of $\gamma$ trials) are given by the following two Equations \ref{BinE} and \ref{BinV} \cite[]{Bezak2014}:

\begin{equation}
\mathbb{E}\left[N\right] = \lambda_u = p\, \gamma
\label{BinE}
\end{equation}

\begin{equation}
\operatorname{Var}\left[N\right] = V_u = p(1-p)\, \gamma
\label{BinV}
\end{equation}

Note that $\operatorname{Var}\left[N\right]<\mathbb{E}\left[N\right]$ for any $\gamma>0$, $0<p<1$, which is a characteristic property of the Binomial distribution. By rearranging Equations \ref{BinE}  and \ref{BinV}, we obtain the expressions for $p$ and $\gamma$ given in Equations \ref{BinP}  and \ref{BinG}:

\begin{equation}
p=\frac{\lambda_u-V_u}{\lambda_u}
\label{BinP}
\end{equation}

\begin{equation}
\gamma=-\frac{\lambda_u^2}{V_u-\lambda_u}
\label{BinG}
\end{equation}

Equations \ref{BinP} and \ref{BinG} can be used to calibrate the count model to the series of the numbers of exceedances of threshold $u$, by using the sample mean and sample variance of excess counts. It is interesting to note that $\gamma$ has the same expression for $r$ in the Negative Binomial (Equation \ref{NegBinR}), with a minus sign, adjusting for the fact that $V_p-\lambda_p$ is negative in the case of under-dispersion.

We want to characterize how the variance of the number of exceedances changes for an arbitrary threshold. Let $V$ and $\lambda$ denote the variance and the expected value of the number of exceedances, respectively, for a generic threshold. By rearranging Equation \ref{BinG} and considering that $\gamma$ does not depend on the threshold (find a demonstration of this in Appendix G), we obtain the expression for $V$ given in Equation \ref{BinV_lambda}:

\begin{equation}
\begin{split}
V = \lambda - \frac{\lambda^2}{\gamma} = \lambda \left(1-\frac{\lambda}{\gamma}\right) 
\label{BinV_lambda}
\end{split}
\end{equation}

The dispersion parameter $1/\gamma>0$ in Equation \ref{BinV_lambda} controls how much smaller the variance $V$ is relative to the mean number of exceedances $\lambda$, through the subtractive term $-\lambda^2/\gamma$. Notably, $V$ tends toward $\lambda$ (i.e., equi-dispersion) as $\gamma\rightarrow+\infty$.

We can now derive the return period $T$ for processes exhibiting under-dispersion. From the expression of the Binomial distribution (Equation \ref{Bin}) and considering Equation \ref{BinP}, the probability of no events in a given year $t$—i.e., in the interval [$t-1,~ t$)—, and the probability of at least one event in that year are given by Equations \ref{pX_t_B0} and \ref{pX_t_B1}, respectively:

\begin{equation}
P \left(\left[ N_{t}-N_{t-1} \right]=0 \right) = 
(1-p)^\gamma= 
\left( \frac{\lambda}{V} \right)^{-\gamma} =
\left(1-\lambda/\gamma\right)^{\gamma}
\label{pX_t_B0}
\end{equation}

\begin{equation}
P \left(\left[ N_{t}-N_{t-1} \right]>0 \right) =
1-P(\left[N_{t}-N_{t-1}\right]=0)=
1- \left( \frac{\lambda}{V} \right)^{-\gamma} =
1-\left(1-\lambda/\gamma\right)^{\gamma}
\label{pX_t_B1}
\end{equation}

Note that the right-hand side of Equations \ref{pX_t_B0} and \ref{pX_t_B1} was obtained by replacing Equation \ref{BinV_lambda}. By definition, the return period is equal to the inverse of the probability of observing an annual maximum (Equation \ref{EqT}); by substituting Equation \ref{pX_t_B1} into Equation \ref{EqT}, we obtain the return period for processes exhibiting under-dispersion, given in Equation \ref{EqT_B}:

\begin{equation}
\begin{split}
T=\frac{1}{1- \left(\lambda /V \right)^{-\gamma}} = \\ =
\frac{1}{1- \left(1-\lambda/\gamma\right)^{\gamma}}
\label{EqT_B}
\end{split}
\end{equation}

Notably, the limit of this expression for $T$ as $V \rightarrow \lambda$ matches the return period derived for a homogeneous Poisson process (i.e., under equi-dispersion; see Equation \ref{EqT}), as shown in Appendix E (Equation \ref{EqLim}). This suggests that Equation \ref{EqT_B} can be viewed as a generalization of Equation \ref{EqT}. The mathematical structure of $T$ for under-dispersed processes shares similarities with the return period for over-dispersed processes (Equation \ref{EqT_NB}). It is straightforward to show that the two formulations are equivalent, as discussed in the main body of this article (see Equation \ref{EqT_Gen_psi}).

\section{Threshold effects on the dispersion of exceedance counts}

\renewcommand{\theequation}{G\arabic{equation}}
\setcounter{equation}{0}

Consider events that occur according to some count model $f_{N_e}(k)=P(N_e=k)$, where $N_e$ represents the number of events in a unit time (e.g., a year, in hydrologic applications). For instance, $f_{N_e}(k)$ could be Poisson, Negative Binomial, or Binomial. We indicate as $\lambda_e$ and $V_e$ the mean and the variance of $N_e$, respectively. 

Now, consider a threshold $Q$ which is exceeded by the generic event with probability $p_Q$. The number of events $N_Q$ exceeding such a threshold within the unit time, conditioned on a fixed $N_e$, will be distributed according to a Binomial with parameters ($N_e$, $p_Q$), as shown in Equation \ref{EqNq_cond_Ne}:

\begin{equation}
\begin{split}
N_Q|\,N_e \sim \binom{N_e}{N_Q}\, \left(p_Q-1 \right) ^{N_e-N_Q}\, p_Q^{N_Q}
\label{EqNq_cond_Ne}
\end{split}
\end{equation}

The expected value $\lambda$ of $N_Q$ can be derived from the Total Expectation theorem as illustrated in Equation \ref{EqTotalExpct}:

\begin{equation}
\begin{split}
\lambda=\mathbb{E}\left[N_Q\right] = \mathbb{E}\left[ \mathbb{E}\left[N_Q|\, N_e\right]\right] =
\lambda_e \, p_Q
\label{EqTotalExpct}
\end{split}
\end{equation}

Note that, to obtain the right-hand side of the above Equation \ref{EqTotalExpct}, we substituted the mean number of successes in a Binomial (Equation \ref{BinE}). In a similar way, we can obtain the variance $V$ of $N_Q$ by applying the Total Variance theorem and considering the mean and variance of the number of successes in a Binomial count process (Equations \ref{BinE} and \ref{BinV}), as illustrated in Equation \ref{EqTotalVar}:

\begin{equation}
\begin{split}
V= \operatorname{Var}\left[N_Q\right] = 
\mathbb{E}\left[\,  \operatorname{Var}\left[N_Q|\, N_e\right] \, \right] +
\operatorname{Var} \left[ \, \mathbb{E}\left[N_Q|\, N_e\right] \, \right] = \\ =
\mathbb{E}\left[\,  p_Q\left(1-p_Q\right)N_e \, \right] +
\operatorname{Var} \left[ \, N_e \, p_Q \, \right] = \\ =
p_Q\left(1-p_Q\right)\, \mathbb{E}\left[\,  N_e \, \right] +
p_Q^2 \, \operatorname{Var} \left[ \, N_e \, \right] = \\ =
p_Q\left(1-p_Q\right)\, \lambda_e +
p_Q^2 \, V_e 
\label{EqTotalVar}
\end{split}
\end{equation}

We can now calculate the dispersion ratio and then the inverse of the parameters $r$ and $\gamma$ of the Negative Binomial and Binomial count models, respectively (Equations \ref{NegBinR} and \ref{BinG}), as shown in Equations \ref{disp_ratio}, \ref{disp_r}, and \ref{disp_gamma}:

\begin{equation}
\begin{split}
\frac{V}{\lambda} = 
\frac
{p_Q\left(1-p_Q\right)\, \lambda_e + p_Q^2 \, V_e }
{\lambda_e \, p_Q} = \\ =
\frac{V_e}{\lambda_e}\, p_Q + 1-p_Q = \\ =
\left(\frac{V_e}{\lambda_e}-1\right)\, p_Q + 1
\label{disp_ratio}
\end{split}
\end{equation}

\begin{equation}
\begin{split}
\frac{1}{r}= \frac{1}{\lambda}\left(\frac{V}{\lambda}-1\right) = \\ = 
\frac{1}{\lambda_e \, p_Q}
\left(\frac{V_e}{\lambda_e}\, p_Q + 1-p_Q-1\right) = \\ =
\frac{1}{\lambda_e}\, \left( \frac{V_e}{\lambda_e}-1\right)
\label{disp_r}
\end{split}
\end{equation}

\begin{equation}
\begin{split}
\frac{1}{\gamma}= -\frac{1}{\lambda}\left(\frac{V}{\lambda}-1\right) = \\ = 
-\frac{1}{\lambda_e \, p_Q}
\left(\frac{V_e}{\lambda_e}\, p_Q + 1-p_Q-1\right) = \\ =
\frac{1}{\lambda_e}\, \left( 1-\frac{V_e}{\lambda_e} \right)
\label{disp_gamma}
\end{split}
\end{equation}


Equations \ref{disp_ratio}, \ref{disp_r}, and \ref{disp_gamma}, providing the dispersion ratio $V/\lambda$, and the inverse of $r$ and $\gamma$, respectively, all obtained for the mean $\lambda$ and the variance $V$ of the exceedance counts associated with an arbitrary threshold $Q$, reveal that these quantities are functions only of the expected value $\lambda_e$ and variance $V_e$ of the original process described by $f_{N_e}$, prior of the definition of any threshold. In other words $V/\lambda$, $r$ and $\gamma$—which are all measures of the degree of dispersion—do not depend on the selected threshold value and remain constant for varying thresholds. 

This result is used in this work to express $V$ as a function of $\lambda$, regardless of the threshold (Equations \ref{NegBinV_lambda} and \ref{BinV_lambda}), and extend the expression for the return period to processes exhibiting over- or under-dispersion in the exceedance counts (Equations \ref{EqT_NB}, \ref{EqT_B}, and \ref{EqT_Gen}, gathering together the former two).

\section*{Acknowledgments}
\addcontentsline{toc}{section}{Acknowledgments}
The authors wish to thank the three reviewers of the article by \cite{Dellaira2025LangComm}, including Prof. Ataur Rahman, for their insightful comments, which highlighted the need for more general formulations of Langbein's equation. Their constructive feedback encouraged the development of the present work.

\bibliographystyle{apalike}   
\bibliography{references}  

\end{document}